\documentclass[aps, prb, superscriptaddress, reprint]{revtex4-2}
\usepackage{amsmath}
\usepackage{bbold}
\usepackage{mathdots}
\usepackage{color}
\usepackage{bm,graphicx,hyperref}

\hypersetup{%
  breaklinks = {true},
  citecolor = {blue},
  colorlinks = {true},
  linkcolor = {red},
}

\begin{document}

\title{Microscopic theory of ionic motion in solids}

\author{Aleksandr Rodin}
\affiliation{Yale-NUS College, 16 College Avenue West, 138527, Singapore}
\affiliation{Centre for Advanced 2D Materials, National University of Singapore, 117546}

\author{Keian Noori}
\affiliation{Centre for Advanced 2D Materials, National University of Singapore, 117546}
\affiliation{Institute for Functional Intelligent Materials, National University of Singapore, 117544}

\author{Alexandra Carvalho}
\affiliation{Centre for Advanced 2D Materials, National University of Singapore, 117546}
\affiliation{Institute for Functional Intelligent Materials, National University of Singapore, 117544}

\author{and A. H. Castro Neto}
\affiliation{Centre for Advanced 2D Materials, National University of Singapore, 117546}
\affiliation{Institute for Functional Intelligent Materials, National University of Singapore, 117544}
\affiliation{Department of Materials Science Engineering, National University of Singapore, 117575}
    
\date{\today}

\begin{abstract}

Drag and diffusion of mobile ions in solids are of interest for both purely theoretical  and applied scientific communities. This article proposes a theoretical description of ion drag in solids that can be used to estimate ionic conductivities in crystals, and forms a basis for the rational design of solid electrolyte materials. Starting with a general solid-state Hamiltonian, we employ the non-equilibrium path integral formalism to develop a microscopic theory of ionic transport in solids in the presence of thermal fluctuations. As required by the fluctuation-dissipation theorem, we obtain a relation between the variance of the random force and friction. Because of the crystalline nature of the system, however, the two quantities are tensorial. We use the drag tensor to write down the formula for ionic mobility, determined by the potential profile generated by the crystal's ions.

\end{abstract}  

\maketitle

\section{Introduction}
\label{sec:Introduction}
As a part of the search for improved energy storage methods~\citep{IEA2020, Yang2018}, substantial attention has been dedicated to the study and development of solid-state batteries in the last decade.~\cite{Bachman2016, Manthiram2017, Famprikis2019} This technology relies on the use of solid electrolytes to conduct ions between the anode and the cathode. The use of all-solid components is advantageous from the safety point of view due to the increased stability of solid-solid interfaces compared to solid-liquid interfaces~\citep{Wang2019}. The main technological challenge lies in finding solid electrolyte materials with a high ionic conductivity at room temperature.

Four main characteristics distinguish solid electrolytes from their liquid counterparts. First, unlike liquid electrolytes, which act as sources of reagents in addition to providing a pathway between the electrodes, solid electrolytes act exclusively as bridges connecting the electrodes and are not consumed in the process of operation. Second, the solid framework through which the ions flow is not mobile, although its atoms vibrate around their equilibrium positions. Because of the periodicity of the vibrational motion, the interaction between the mobile ions and the framework cannot be generally regarded as a collection of uncorrelated collisions, as would be the case in a liquid. Hence, it is not immediately obvious that treating the motion of the mobile ions as Brownian is appropriate, suggesting that the Nernst-Einstein relation might be inapplicable in this case~\citep{Wang2015,Marcolongo2017}. The third aspect that sets solid electrolytes apart is a non-trivial potential landscape produced by the framework ions and electrons, through which the mobile ions navigate. This landscape contains local energy minima, which can function as traps for the mobile ions, requiring them to regularly overcome potential barriers of fractions of electronvolts during their motion.~\cite{Kanno2001, Kuhn2013, Bron2014, Kuhn2014, Seino2014, Wang2015, He2017, Muy2018, DiStefano2019} The energy needed to escape the local minima originates from the framework itself as the thermally vibrating lattice kicks the mobile ions. Finally, the fourth key difference lies in the role played by quantum mechanics. Although the heavy ions traveling through the framework at typical battery operation temperatures are classical objects, they interact strongly with the quantum electrons of the framework. Moreover, the vibrational modes of the framework are also quantum mechanical objects with Bose statistics. These distinguishing features indicate that the problem of ionic conductors falls in the domain of solid-state physics and should be addressed in this context.

On the theoretical side, {\color{black}nudged elastic band (NEB) calculations have provided insight into low-energy pathways for mobile ions in solid electrolytes~\citep{Wang2015}. Classical molecular dynamics (MD) and \textit{ab initio} molecular dynamics (AIMD) simulations, meanwhile, are an integral part of research in ionic conductors}. They have been instrumental in shedding light on the atomic-scale processes behind the ionic conduction by identifying body-centered cubic crystals as the optimal lattice structures for fast ionic conduction~\citep{Wang2015, He2017}, demonstrating the importance of cooperative (correlated) ionic transport~\cite{Deng2015, He2017}, studying the role of frustration mechanisms~\citep{Kozinsky2016, Adelstein2016, DiStefano2019}, exploring the effects of anharmonicity~\citep{Brenner2020, Ding2020}, and providing a deeper understanding of the role played by the lattice dynamics~\citep{Krauskopf2017, Muy2018} and structural modification.~\citep{DeKlerk2016, Deng2017} {\color{black} AIMD simulations have also been used, with some success, to calculate the conductivity of mobile ions in solid electrolytes via computation of the tracer diffusion coefficient, $D_{tr}$, and its insertion into the Nernst-Einstein equation~\citep{Yang2015,Wang2015,Marcolongo2017,mo2012first,miwa2021molecular,he2018statistical,wang2019lithium}.  The Nernst-Einstein equation, however, has been shown to be invalid in the presence of correlation between particles, leading to an underestimation of conductivity unless these correlations are accounted for~\citep{Wang2015,Marcolongo2017,pang2021mechanical}}.

Despite MD simulations' undisputed success and utility, the technique has some limitations, mainly originating from the computational cost. To ensure accuracy, the time steps in the simulations must be small (on the order of femtoseconds), meaning that the total simulation time is often limited to a few nanoseconds. Consequently, it is not uncommon to use temperatures much higher than those expected in device operation {\color{black}(of, e.g., solid state batteries) to speed up the dynamics and observe sufficient ionic activity within the limited time window\cite{qi2021bridging,he2018statistical}. Even then, the computation times are substantially shorter than experimentally relevant time scales. Moreover, given the structural complexity of many ionic conductors, simulations have generally been restricted to simple systems, or more complex systems limited to a few unit cells since increasing the system size renders the calculations prohibitively expensive. Accordingly, the study of multiple ions separated by large distances is highly challenging. It is therefore apparent that a complementary theoretical methodology, able to address the interplay between ions and the lattice over greater time and length scales, is desirable.}

\begin{figure}
    \centering
    \includegraphics[width=8cm]{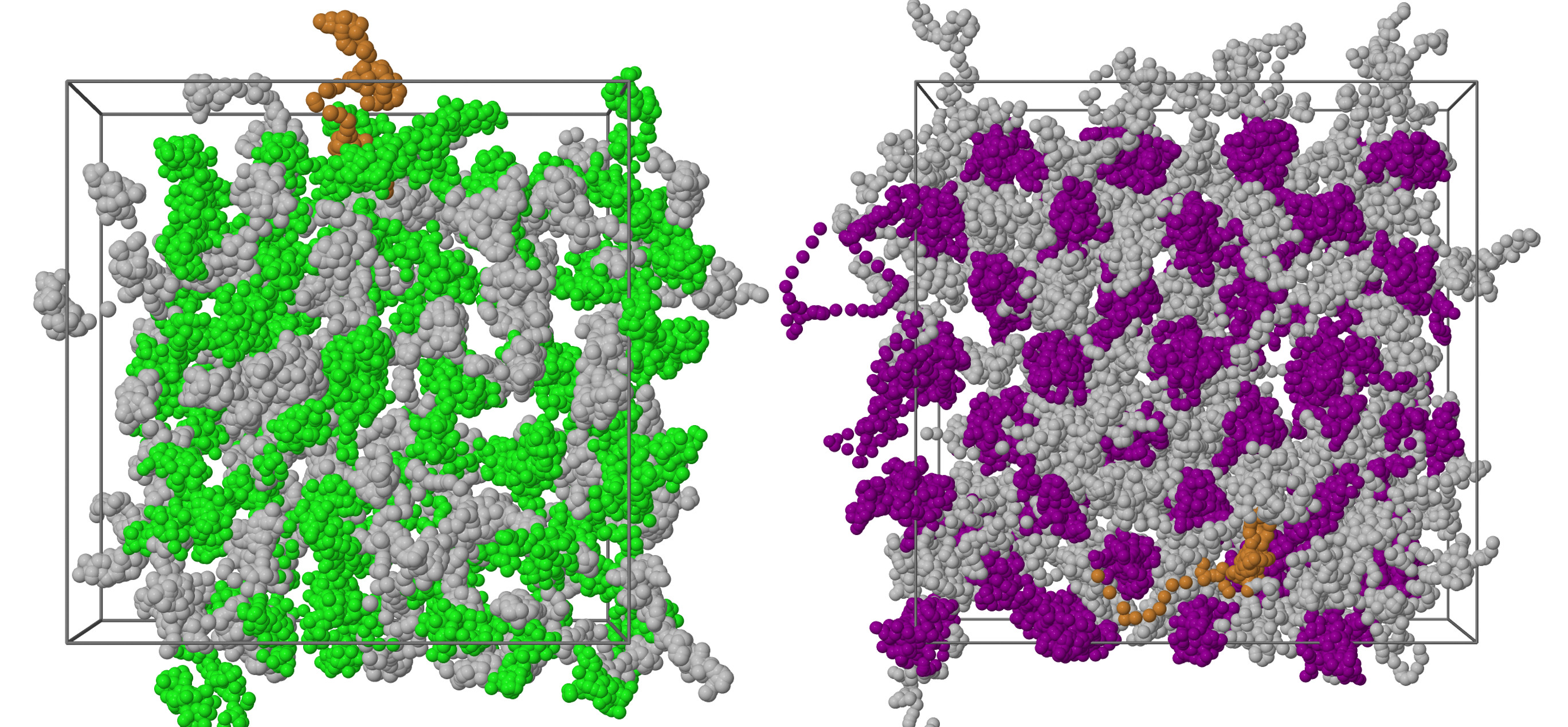}
    \caption{Diffusion trajectories of ions in AgCl at 600 K (left panel) and $\alpha$-AgI at 700~K (right panel), obtained from \textit{ab initio} molecular dynamics simulations. Ag atoms are represented in grey, while Cl and I atoms are represented in green and purple, respectively. The trajectory of a single Ag ion is highlighted in orange. The positions are represented every 0.1~ps, for a total time of 10~ps.}
    \label{fig:trajectories}
\end{figure}
As mentioned above, lattice vibrations impart kinetic energy onto mobile ions, allowing them to escape potential-energy valleys.
To illustrate this, consider the examples of $\alpha$-AgI and AgCl.
The first is a superionic conductor, where, at any given time, a large fraction of the silver ions are
mobile, equivalently to mobile interstitials.\cite{liou.PhysRevB.41.10481}
In contrast, AgCl is a solid with rock-salt structure, where mobile Ag ions are thermally generated as the interstitial moieties of Frenkel pairs~\cite{friauf1977determination}.
These mobile ions travelling through the solid have to regularly escape local potential minima assisted by the framework's thermal fluctuations. This motion resembles a ``hopping" transport, where the ions oscillate around a local minimum before moving to an adjacent one. This is evident in the trajectories obtained from molecular dynamics simulations of the thermal diffusion in the ionic conductors AgCl and $\alpha$-AgI (Fig.~\ref{fig:trajectories}). 

The diffusion of the mobile ions is reminiscent of a random walk associated with Brownian motion. Just as in the case of Brownian motion, however, the fluctuation-dissipation theorem demands that the lattice-to-ions energy flow must be accompanied by the reverse process, where the lattice saps the energy from the moving ions similar to the macroscopic drag phenomenon. Note that, unlike the traditional drag and diffusion in liquids, the size of the moving particles (mobile ions) is comparable to that of the bath particles (lattice ions). Therefore, each collision between the two components can substantially modify the energy of the mobile ions. By contrast, Brownian particles (pollen organelles in the original experiment) experience an astronomical number of collisions before moving by an appreciable amount, allowing one to treat the collisions as uncorrelated white noise.

The problem of a small mobile particle coupled to a dissipative thermal bath has been of interest to the physics community for a long time.~\citep{Feynman1963, Caldeira1981} In recent years, there have been significant advances in understanding the dynamics of impurities immersed in bosonic~\citep{Caldeira1995, Schecter2012, Peotta2013, Dehkharghani2015, Petkovic2016} and fermionic~\citep{Caldeira1995, CastroNeto1996, Pasek2019} systems. The authors of Ref.~\citep{Lampo2017} demonstrated the emergence of the Brownian motion in $D$-dimensional Bose-Einstein condensate systems while Ref.~\cite{Petkovic2020} focused on the microscopic origins of friction in one-dimensional quantum liquids. Because these problems are commonly viewed from the perspective of (ultra-) cold atom experiments, they are typically formulated in one dimension.

In this work, we construct a general microscopic theory applicable to three dimensions to describe the motion of ions through a solid framework that can be used to estimate ionic conductivities and form a basis for the design of solid electrolyte materials. {\color{black}It will be shown that our approach results in a simple and intuitive temperature-free expression for the steady-state ionic mobility in a crystal, thereby mitigating the two predominant obstacles associated with MD simulations. In addition, we demonstrate the practical application of our formalism using first principles calculations to compute approximate ionic mobilities for a range of small crystals, laying a promising foundation on which further refinements may be developed.} (Note that the theoretical formalism is developed in Sections~\ref{sec:Effective_Hamiltonian}--~\ref{sec:Drift_Diffusion}; readers only interested in numerical applications can skip directly to Section~\ref{sec:Numerical_Results}.) In Sec.~\ref{sec:Effective_Hamiltonian}, we set up the Hamiltonian for a system with vibrational modes and mobile masses. We also demonstrate how the motion of the mobile particles can be calculated using the classical framework. Section~\ref{sec:System_Dynamics} focuses on the derivation of the semiclassical equations of motion for the mobile particles starting from the non-equilibrium quantum formulation. In Sec.~\ref{sec:Drift_Diffusion}, we establish the fluctuation-dissipation relation in crystalline materials and derive the expression of the ionic mobility in solids. A prototypical application of our formalism to the determination of ionic mobilities in real crystal, by way of \textit{ab initio} numerical calculations, in given in Sec.~\ref{sec:Numerical_Results}. Conclusions are found in Sec.~\ref{sec:Conclusions}.

\section{Effective Hamiltonian}
\label{sec:Effective_Hamiltonian}

{\color{black}The aim of this section is to set up the effective Hamiltonian which will be used to study the system dynamics. In Sec.~\ref{sec:Ionic_Hamiltonian}, we start with a general solid state Hamiltonian and integrate out the electronic degrees of freedom to write down an effective Hamiltonian that depends only on the ionic coordinates. Next, in Sec.~\ref{sec:Ion-framework}, we split the system's ions into two groups: the stationary framework and the mobile species. In Sec.~\ref{sec:Classical}, we use the classical approach to integrate out the framework degrees of freedom and obtain the equations of motion for the mobile ions with a memory term and a stochastic thermal component. The result of this section will be used in Sec~\ref{sec:Drift_Diffusion} to study the drift and diffusion of mobile ions in a solid framework. Section~\ref{sec:Quant} reformulates the Hamiltonian from Sec.~\ref{sec:Ion-framework} using quantum mechanics to be used in the path integral derivation in Sec.~\ref{sec:System_Dynamics}. We will show that in the semi-classical limit, the path integral approach gives the same result as the classical formulation. Therefore, the readers who are interested only in this limit can skip to Sec.~\ref{sec:Drift_Diffusion} directly after Sec.~\ref{sec:Classical}.
}

\subsection{Ionic Hamiltonian}
\label{sec:Ionic_Hamiltonian}

The most general microscopic Hamiltonian for a solid composed of ions and electrons can be written as 

\begin{equation}
    H = K_e + V_{ee} + K_i + V_{ii} + V_{ei} + E_e + E_i\,,
    \label{eqn:Microscopic_H}
\end{equation}
where $K_e$ ($K_i$) is the kinetic energy of electrons (ions), $V_{ee}$ ($V_{ii}$) is the electron-electron (ion-ion) interaction, $V_{ei}$ is the electron-ion interaction, and $E_e$ ($E_i$) is the external potential acting on electrons (ions). Generally, the external potentials $E_{e/i}$ can be time-dependent, resulting in a non-equilibrium behavior. 

From the practical standpoint, the time variation of $E_e(t)$ seen in applications is expected to be sufficiently slow to treat its impact on the electrons quasistatically. Additionally, because ions are much heavier than electrons, one can follow the Born-Oppenheimer approximation and view them as static, as far as the electrons are concerned. Consequently, we can write the electronic Hamiltonian operator as 
\begin{equation}
    \hat{H}_e\left(\{\mathbf{R}\},t\right) = \hat{K}_e + \hat{V}_{ee} + \hat{V}_{ei}\left(\left\{\mathbf{R}\right\}\right) + \hat{E}_e(t)\,,
    \label{eqn:Electronic_H}
\end{equation}
where $\{\mathbf{R}\}$ is the set of all the ionic coordinates. We stress that $\{\mathbf{R}\}$ and $t$ are parameters of the electronic Hamiltonian operator, not dynamic variables.

It is useful to write $\hat{V}_{ei}\left(\left\{\mathbf{R}\right\}\right) = \hat{V}_{ei}\left(\left\{\mathbf{R}^0\right\}\right) + \delta \hat{V}_{ei}\left(\left\{\mathbf{R}\right\}\right)$, where $\hat{V}_{ei}\left(\left\{\mathbf{R}^0\right\}\right)$ is the interaction between electrons and the system's native ions when the ions are located at their energy minima. Note that $\delta \hat{V}_{ei}\left(\left\{\mathbf{R}\right\}\right)$ can also include the interaction of the system's electrons with extra ions added to the system. We also define $\hat{H}_e^0 = \hat{K}_e + \hat{V}_{ee} + \hat{V}_{ei}\left(\left\{\mathbf{R}^0\right\}\right)$ as the electronic Hamiltonian in a stationary unperturbed solid, so that the full electronic Hamiltonian is $\hat{H}_e = \hat{H}_e^0 + \delta \hat{V}_{ei}\left(\left\{\mathbf{R}\right\}\right) + \hat{E}_e(t)$.

Although the composition of solid electrolytes can vary widely, they must be electronic insulators to guarantee that the current passing through them is exclusively ionic. Consequently, $\hat{H}_e^0$ must possess a sufficiently wide gap for $ \delta \hat{V}_{ei}\left(\left\{\mathbf{R}\right\}\right) + \hat{E}_e(t)$ not to create electron-hole excitations leading to electronic transport. As a result, the perturbation only leads to a modification of $\hat{H}_e^0$'s energies, following the adiabatic theorem. 
Formally, the Helmholtz free energy for the electrons is given by

\begin{align}
    F_e =&
    -
  \frac{1}{\beta}\sum_n \ln  \left|-\beta \left[G_n^{-1} - \delta V_{ei}\left(\left\{\mathbf{R}\right\}\right) -E_e(t)\right]\right|
    \nonumber
    \\
    =&
   \underbrace{ -
  \frac{1}{\beta}\sum_n \ln  \left|-\beta G_n^{-1}\right|}_{F_e^0}
    \nonumber
    \\
    -& \frac{1}{\beta}\sum_n
    \ln \left|1 - G_n\left[\delta V_{ei}\left(\left\{\mathbf{R}\right\}\right) +E_e(t)\right]\right|
   \,,
   \label{eqn:F_e_BO}
\end{align}
where $\beta^{-1} = k_BT$, $k_B$ is the Boltzmann constant, $T$ is the temperature, $G^{-1}_n = i\omega_n + \mu - H^0_e$ is the Green's function matrix, $\mu$ is the chemical potential, and $\omega_n$ are the fermionic Matsubara frequencies. $F_e^0$ is the electronic contribution to the free energy in an unperturbed system and the second term gives the perturbation-induced correction. The latter can be rewritten as

\begin{align}
    \delta F_e =
    -& \frac{1}{\beta}\sum_n
    \ln\left|1 - G_n\delta V_{ei}\left(\left\{\mathbf{R}\right\}\right) \right|
    \nonumber
    \\
    -&
    \frac{1}{\beta}\sum_n
    \ln\left|1 - \left[1 - G_n\delta V_{ei}\left(\left\{\mathbf{R}\right\}\right)\right]^{-1}G_n E_e (t)\right|
   \,.
   \label{eqn:delta_F_e}
\end{align}

Combining the electronic free enegy with the remaining terms of $H$ gives the effective Hamiltonian describing the ionic motion:

\begin{align}
    H_i =& K_i 
    +
    \overbrace{V_{ii} +F_e^0 - \frac{1}{\beta}\sum_n
    \ln\left|1 - G_n\delta V_{ei}(\{\mathbf{R}\}) \right|}^{U(\{\mathbf{R}\})}
     + E_i(t)
     \nonumber
     \\
    -&\frac{1}{\beta}\sum_n
    \ln\left|1 - \left[1 - G_n\delta V_{ei}\left(\left\{\mathbf{R}\right\}\right)\right]^{-1}G_n E_e (t)\right|
     \,.
     \label{eqn:H_i}
\end{align}
$U(\{\mathbf{R}\})$ describes the interaction between ions, including the electronic effects, and can be computed \textit{ab initio} using density functional theory (DFT) by calculating the energy of a system with ionic coordinates $\left\{\mathbf{R}\right\}$.

The last term in Eq.~\eqref{eqn:H_i} gives the energy due to the electrons' interaction with the external potential, including the effects of the perturbed ionic background. For a stable solid not undergoing a phase transition, it is reasonable to expect that the system-wide electronic density will not change drastically in response to the small shift in ionic coordinates around their equilibrium positions. Therefore, we drop $\delta V_{ei}\left(\left\{\mathbf{R}\right\}\right)$ in this expression, rendering it independent of the ionic position. Consequently, the effective ionic Hamiltonian involves only the first line of Eq.~\eqref{eqn:H_i} because the last line does not depend on $\mathbf{R}$ after $\delta V_{ei}$ is dropped and, to the leading order in $E_e$, gives the Hartree energy of electrons in an external potential.

\subsection{Ion-framework interaction}
\label{sec:Ion-framework}

To study the ionic motion described by Eq.~\eqref{eqn:H_i}, we divide the ions into two groups: those that propagate through the solid (mobile ions) and those that vibrate around their equilibrium positions and provide the solid framework (stationary ions). We make this distinction explicit by rewriting Eq.~\eqref{eqn:H_i} as

\begin{equation}
    H_i = K_i^M +  K_i^S + U\left(\left\{\mathbf{r}\right\},\left\{\mathbf{u}\right\}\right) + E_i(t)\,,
    \label{eqn:H_i_separated}
\end{equation}
where $\mathbf{r}$ ($\mathbf{u}$) are the positions of the mobile (stationary) ions.

At this point, there are two main approaches that can be used to solve the problem. On the one hand, it is possible to view the system as entirely classical to obtain the trajectories of the mobile ions. The benefit of this method is that it is conceptually simpler and puts fewer restrictions on the interaction between the mobile and the stationary ions. The downside is that the thermal motion of the framework is not automatically captured from the partition function. In addition, the classical formalism might be inapplicable when quantum effects  become important, such as for proton diffusion or in certain cold-atom setups.

On the other hand, one can start by assuming that the framework ions do not deviate substantially from their equilibrium positions $\mathbf{u}^0$ and expand the potential energy term for small displacement $\boldsymbol{\delta} = \mathbf{u} -  \mathbf{u}^0$. Following this expansion, one writes $\boldsymbol{\delta}$ in terms of the oscillatory modes of the framework. The interaction between the two groups of ions then becomes linear in $\boldsymbol{\delta}$. By second-quantizing the modes, quantum mechanics is included in the problem formulation, delaying the semi-classical treatment until the very end. This approach makes it possible to include quantum-mechanical corrections beyond the leading-order classical behavior. Most importantly, this method explicitly encodes the thermal occupancy of phonons, producing the correct fluctuation-dissipation relation. Naturally, in the classical limit, the two approaches should give identical results. Therefore, for the sake of completeness, we show both treatments.

\subsection{Classical Formulation}
\label{sec:Classical}

It is convenient to start by separating the interaction term into three components: $U\left(\left\{\mathbf{r}\right\},\left\{\mathbf{u}\right\}\right) \rightarrow U^S\left(\left\{\mathbf{u}\right\}\right) + U^M\left(\left\{\mathbf{r}\right\}\right) + U\left(\left\{\mathbf{r}\right\},\left\{\mathbf{u}\right\}\right)$. Next, suppressing the function arguments for brevity, we can write the Lagrangian for the system as

\begin{equation}
    L = \left(K_i^M - U^M\right) + \left(K_i^S - U^S\right)  -U - E^S_i - E^M_i \,.
    \label{eqn:Lagrangian}
\end{equation}

Assuming that the motion of the framework ions can be described using the harmonic approximation, the homogeneous portion of the framework Lagrangian becomes

\begin{equation}
    K_i^S - U^S \rightarrow 
     \frac{1}{2}\dot{\mathbf{u}}^T\mathbf{m}\dot{\mathbf{u}}
    -
    \frac{1}{2}\mathbf{u}^T \mathbf{V} \mathbf{u}\,.
    \label{eqn:Framework_Lagrangian}
\end{equation}
Here, we combined the positions of the framework ions into a $DI$-dimensional vector $\mathbf{u} = \bigoplus_{j=1}^I \mathbf{u}_j$, where $I$ is the number of the framework ions and $D$ is the system dimensionality. $\mathbf{m} = \bigoplus_{j=1}^I m_j\mathbf{1}_{D\times D}$ is a block-diagonal matrix, where $m_j$ is the mass of the $j$th framework ion.

The homogeneous equation of motion $\mathbf{m}\ddot{\mathbf{u}} =  -\mathbf{V}\mathbf{u}$ can be transformed into a symmetric eigenvalue problem by first defining $\tilde{\mathbf{u}} = \mathbf{m}^{1/2}\mathbf{u}$ so that

\begin{equation}
    \ddot{\tilde{\mathbf{u}} }
    = 
    -\Omega_s^2\tilde{\mathbf{u}}
    = 
    -
    \mathbf{m}^{-1/2}\mathbf{V}\mathbf{m}^{-1/2}\tilde{\mathbf{u}} 
    =
    -
   \tilde{\mathbf{V}}\tilde{\mathbf{u}} 
    \,,
    \label{eqn:u_EOM}
\end{equation}
with the eigenvectors $\varepsilon_s$ and corresponding eigenvalues $\Omega_s$. Hence, we can write $\tilde{\mathbf{u}}(t) = \boldsymbol{\varepsilon}\zeta(t)$ [so that $\mathbf{u}(t) =\mathbf{m}^{-1/2} \boldsymbol{\varepsilon}\zeta(t)$], where $\zeta(t)$ is a column vector of normal coordinates giving the amplitude of each mode, while $\boldsymbol{\varepsilon}$ is a row of column vectors $\varepsilon_s$.

Returning to the inhomogeneous equation of motion for the framework ions, we write

\begin{align}
    &\mathbf{m}\ddot{\mathbf{u}} = -\mathbf{V}\mathbf{u} - \nabla_\mathbf{u}\left(U + E_i^S\right)
    \nonumber
    \\
    \rightarrow 
    &\ddot{\tilde{\mathbf{u}}} = -\tilde{\mathbf{V}}\tilde{\mathbf{u}} - \mathbf{m}^{-1/2}\nabla_\mathbf{u}\left(U + E_i^S\right)
    \nonumber
    \\
    \rightarrow
    &\ddot{\zeta} = -\boldsymbol{\Omega}^2\zeta - \boldsymbol{\varepsilon}^{-1}\mathbf{m}^{-1/2}\nabla_\mathbf{u}\left(U + E_i^S\right)\,,
    \label{eqn:zeta_EOM}
\end{align}
where $\boldsymbol{\Omega}^2 = \boldsymbol{\varepsilon}^{-1}\tilde{\mathbf{V}}\boldsymbol{\varepsilon}$ is a diagonal matrix of the squared eigenfrequencies.

For a single normal coordinate, the expression above takes the form $\ddot{\zeta}_j = - \Omega_j^2 - f_j$, which can be solved using the Green's functions. Recalling that the Green's function for a harmonic oscillator is given by

\begin{equation}
     G_j(t,t')= \frac{\sin\left[\Omega_j(t - t')\right]}{\Omega_j}\Theta(t - t')\,,
\label{eqn:Greens_fn}
\end{equation}
we have

\begin{widetext}
\begin{align}
    \zeta_j(t) &= \zeta_j^H(t) - \int^t dt'\frac{\sin\left[\Omega_j(t - t')\right]}{\Omega_j}
   f_j = \zeta_j^H(t) - \int^t dt'\frac{\sin\left[\Omega_j(t - t')\right]}{\Omega_j}
   \left[\boldsymbol{\varepsilon}^{-1}\mathbf{m}^{-1/2}\nabla_\mathbf{u}\left(U + E_i^S\right)\right]_j 
   \nonumber
   \\
    & = \zeta_j^H(t) - \int^t dt'\frac{\sin\left[\Omega_j(t - t')\right]}{\Omega_j}
   \left[\nabla_\mathbf{u}\left(U + E_i^S\right)\right]^T\mathbf{m}^{-1/2}\varepsilon_j \,,
\end{align}
\end{widetext}
where $\zeta_j^H(t)$ is the homogeneous solution and the subscript $j$ at the brackets indicates that we pick out the $j$th element of the column vector. The last line follows from the fact that $\boldsymbol{\varepsilon}$ is an orthogonal matrix, $\mathbf{m}$ is a diagonal matrix, and that the transpose of the expression in the brackets it the expression itself.

Finally, using $\tilde{\mathbf{u}} = \sum_j \zeta_j \varepsilon_j$, we obtain

\begin{align}
    \mathbf{u}(t)
    &=
    \mathbf{u}^0
    +
    \boldsymbol{\delta}^H(t)
    \nonumber
    \\
    &- 
   \sum_j   \mathbf{m}^{-1/2}\varepsilon_j\Bigg\{\int^t dt'\frac{\sin\left[\Omega_j(t - t')\right]}{\Omega_j}
   \nonumber
   \\
   &\times
   \left[\nabla_\mathbf{u}\left(U + E_i^S\right)\right]^T\mathbf{m}^{-1/2}\varepsilon_j\Bigg\}\,,
   \label{eqn:u_Classical}
\end{align}
where $\mathbf{u}^0$ gives the equilibrium positions of the framework ions and $\boldsymbol{\delta}^H(t)$ is the displacement from the equilibrium coming from the homogeneous solution.

Reinserting the expression of $\mathbf{u}(t)$ into the interaction energy $U(\mathbf{u}, \mathbf{r})$, we can calculate the force that this interaction exerts on the mobile ions $-\nabla_\mathbf{r} U(\mathbf{u}, \mathbf{r})= -\nabla_\mathbf{r} U(\mathbf{u}^0+\boldsymbol{\delta}, \mathbf{r})$. By assuming that the framework ions do not move far from the equilibrium, we expand the expression in $\boldsymbol{\delta}$ to obtain

\begin{align}
    &-\nabla_\mathbf{r} U(\mathbf{u}^0+\boldsymbol{\delta}, \mathbf{r})
    \nonumber
    \\
    &\approx
    -\nabla_\mathbf{r} U(\mathbf{u}^0, \mathbf{r})
    -\nabla_\mathbf{r} 
    \left[\nabla_{\mathbf{u}^0} U(\mathbf{u}^0, \mathbf{r})\cdot\boldsymbol{\delta}\right]
    \nonumber
    \\
    &\approx
     -\nabla_\mathbf{r} U(\mathbf{u}^0, \mathbf{r})
    -\nabla_\mathbf{r} 
    \left[\nabla_{\mathbf{u}^0} U(\mathbf{u}^0, \mathbf{r})\cdot\boldsymbol{\delta}^H(t)\right]
    \nonumber
    \\
    &+ \frac{2}{\hbar}\sum_j
    \nabla_{\mathbf{r}(t)}
    \Bigg\{
   Y_j(\mathbf{r}(t))
    \int^t dt' \sin\left[\Omega_j(t - t')\right]
    \nonumber
    \\
    &\times
   \left[Y_j(\mathbf{r}(t'))+W_j(t')\right]\Bigg\}\,,
   \label{eqn:Classical_Force}
   \\
    Y_{s}(\mathbf{r}) &=
    \sqrt{\frac{\hbar}{2\Omega_s}}
    \left[\nabla_{\mathbf{u}^0} U\left(\mathbf{r},\mathbf{u}^0\right)
    \right]^T
    \mathbf{m}^{-1/2}
    \boldsymbol{\varepsilon}_{s}\,,
    \label{eqn:Y}
    \\
    W_s(t)&=\sqrt{\frac{\hbar}{2\Omega_s}}
    \left[\nabla_{\mathbf{u}^0} E_i^S\left(\mathbf{u}^0,t\right)
    \right]^T
    \mathbf{m}^{-1/2}
    \boldsymbol{\varepsilon}_{s}\,.
    \label{eqn:W}
\end{align}

We show below that one arrives at the same expression in the semi-classical limit of the quantum-mechanical formulation, except that the quantum-mechanical treatment explicitly gives the temperature dependence of the homogeneous term.

\subsection{Quantum Formulation}
\label{sec:Quant}

For the quantum-mechanical approach, we start by expanding the potential term to the second order in the framework ion displacement $\boldsymbol{\delta} = \mathbf{u} -  \mathbf{u}^0$ to get

\begin{align}
     &U\left(\left\{\mathbf{r}\right\},\mathbf{u}\right)
     \approx 
     U\left(\left\{\mathbf{r}\right\},\mathbf{u}^0\right) 
     + \left[\nabla_{\mathbf{u}^0} U\left(\left\{\mathbf{r}\right\},\mathbf{u}^0\right)\right]^T \boldsymbol{\delta}
     \nonumber
     \\
     +&\frac{1}{2} \boldsymbol{\delta}^T\left[
     \left(\nabla_{\mathbf{u}^0}\otimes \nabla_{\mathbf{u}^0}\right)
     U\left(\left\{\mathbf{r}\right\},\mathbf{u}^0\right)\right]
     \boldsymbol{\delta}\,.
     \label{eqn:U_expanded}
\end{align}
Note that we combined all the framework coordinates and displacements into two vectors: $\mathbf{u} = \bigoplus \mathbf{u}_l$ and $\boldsymbol{\delta} = \bigoplus \boldsymbol{\delta}_l$, as was done in the classical treatment.

The last portion of Eq.~\eqref{eqn:U_expanded} can be identified as the elastic potential energy with the term in the brackets being the matrix of force constants coupling the displacements of the stationary ions. For a fixed $\left\{\mathbf{r}\right\}$, combining this term with the kinetic energy $K_i^S$ gives rise to a collection of oscillatory modes. Strictly speaking, changing $\left\{\mathbf{r}\right\}$ modifies the force-constant matrix and alters the mode frequencies. However, it is reasonable to expect that, for a stable system, moving the mobile ions through the system does not drastically alter the structure of the stationary framework. Consequently, we will assume that the term in the brackets does not depend on $\left\{\mathbf{r}\right\}$, allowing us to promote $H_i$ to the operator status and write
\begin{align}
    \hat{H}_i &=\sum_{s}\hbar\Omega_{s}\left(a^\dagger_{s}a_{s}+\frac{1}{2}\right)+ E_i^S(\hat{\mathbf{u}},t)
    \nonumber
    \\
    &+ \sum_j \frac{\hat{\mathbf{p}}_j^\dagger \hat{\mathbf{p}}_j}{2M_j} 
    +
   U\left(\left\{\hat{\mathbf{r}}\right\},\mathbf{u}^0\right) 
    + E_i^M\left(\left\{\hat{\mathbf{r}}\right\},t\right)
     \nonumber
     \\
     &+ \left[\nabla_{\mathbf{u}^0} U\left(\left\{\hat{\mathbf{r}}\right\},\mathbf{u}^0\right) \right]^T \hat{\boldsymbol{\delta}}
     \,,
     \label{eqn:H_i_operator}
\end{align}
where the first term is the second-quantized Hamiltonian of the oscillatory modes $s$ with frequency $\Omega_s$ independent of $\left\{\mathbf{r}\right\}$, resulting from combining $K_i^S$ with the last term in Eq.~\eqref{eqn:U_expanded}. If the framework is crystalline, these vibrational modes correspond to phonons and the mode label $s$ denotes the phonon branch and momentum. The first term in the second line of Eq.~\eqref{eqn:H_i_operator} is the kinetic energy $K_i^M$ with $\mathbf{p}_j$ corresponding to the momentum of the $j$th mobile ion and $M_j$ to its mass. Note that we split the effects of the external potential into portions corresponding to stationary and mobile ions.

Using the fact that the external perturbation is not expected to vary substantially on the scale of $\boldsymbol{\delta}$, we write
\begin{align}
    &\hat{E}_i^S(\hat{\mathbf{u}},t) \approx E_i^S(\mathbf{u}^0,t) + \left[\nabla_{\mathbf{u}^0}E_i^S(\mathbf{u}^0,t)\right]^T\hat{\boldsymbol{\delta}}\,.
    \label{eqn:E_i_expanded}
\end{align}
The first term does not depend on the ionic displacement and, therefore, does not impact the system's dynamics. Hence, we drop this term from the Hamiltonian.

To describe $\boldsymbol{\delta}$ in terms of the solid's vibrational modes, recall that, in the harmonic approximation, the displacement is
\begin{equation}
    \hat{\boldsymbol{\delta}} =
    \sum_s \left(a_s +a_s^\dagger \right)\sqrt{\frac{\hbar}{2 \Omega_s}}\mathbf{m}^{-1/2}\boldsymbol{\varepsilon}_{s}\,,
    \label{eqn:delta}
\end{equation}
where $\boldsymbol{\varepsilon}_{s}$ is the polarization vector for mode $s$. Using this definition, we obtain
\begin{align}
    \hat{H}_i
    &=
    \sum_{s}\hbar\Omega_{s}\left(a^\dagger_{s}a_{s}+\frac{1}{2}\right)
    \nonumber
    \\
    &
    +
    \hat{\mathbf{p}}^\dagger \frac{\mathbf{M}^{-1}}{2}\hat{\mathbf{p}}
    + 
    \overbrace{U\left(\hat{\mathbf{r}}\right)
    + E_i^M\left(\hat{\mathbf{r}},t\right)}^{U\left(\hat{\mathbf{r}},t \right)}
    \nonumber
    \\
    &+
    \sum_{s}
    \underbrace{
    \left[
    Y_{s}(\hat{\mathbf{r}})+W_s(t)
    \right]}_{C_s(\hat{\mathbf{r}},t)}
    \left(
    a_{s}
    +
    a_{s}^\dagger \right)
    \,.
    \label{eqn:H_i_Compact}
\end{align}
In writing this expression, we combined all mobile ion positions $\left\{\mathbf{r}\right\}$ and momenta $\left\{\mathbf{p}\right\}$ into $DI$-dimensional coordinates $\mathbf{r} = \bigoplus_{j = 1}^I \mathbf{r}_j$ and $\mathbf{p}= \bigoplus_{j = 1}^I \mathbf{p}_j$, where $I$ is the number of mobile ions.

The second line of Eq.~\eqref{eqn:H_i_Compact} describes the mobile ions in the presence of a potential produced by the framework ions at their equilibrium positions [$U(\hat{\mathbf{r}}) \equiv U\left(\hat{\mathbf{r}},\mathbf{u}^0\right)$] and an external time- and position-dependent perturbation $E_i^M\left(\hat{\mathbf{r}},t\right)$. $\mathbf{M}= \bigoplus_{j=1}^I M_j\mathbf{1}_{D\times D}$ is a block-diagonal matrix.

Finally, the last line of Eq.~\eqref{eqn:H_i_Compact} gives the coupling between the oscillatory modes and the mobile ions [$Y_s\left(\hat{\mathbf{r}}\right)$], and the modes and the external potential [$W_s(t)$]. To make the subsequent derivation more compact, we combine the two coupling terms into one, denoted by $C_s(\hat{\mathbf{r}},t)$, as shown by the underbrace.

\section{System Dynamics}
\label{sec:System_Dynamics}

With the effective time-dependent ionic Hamiltonian given by Eq.~\eqref{eqn:H_i_Compact}, we can now address the dynamics of the system. We begin by formulating the problem using the path integral language and then proceed to extract the semiclassical equations of motion for the mobile ions.

\subsection{Path Integral}
\label{sec:Path_Integral}

Recall that if, at $t = 0$, the system is described by a density operator $\hat{\rho}_0$, then the expectation value of some operator $\hat{O}$ at $\tau > t$ is given by

\begin{align}
    \langle\hat{O}\rangle (\tau) &
    = 
    \frac{\mathrm{Tr}[e^{\frac{i\hat{H}_i\tau}{\hbar}}\hat{O}e^{-\frac{i\hat{H}_i\tau}{\hbar}}\hat{\rho}_0]}{\mathrm{Tr}[\hat{\rho}_0]}
    \nonumber
    \\
   & =
    \sum_{n,\mathbf{r}}\frac{\langle \mathbf{r}, n|e^{\frac{i\hat{H}_i\tau}{\hbar}}\hat{O}e^{-\frac{i\hat{H}_i\tau}{\hbar}}\hat{\rho}_0|\mathbf{r},n\rangle}{\mathrm{Tr}[\hat{\rho}_0]}\,.
    \label{eqn:O_tau_general}
\end{align}
To go from the first line to the second one, we used the fact that the trace of the operator can be taken in any complete basis, allowing us to choose $|\mathbf{r},n\rangle = |\mathbf{r}\rangle\otimes|n\rangle$ with $|n\rangle$ enumerating all the Fock states for the vibrational modes and $|\mathbf{r}\rangle$ corresponding to the multi-particle position states. In this study, we are primarily interested in how the mobile ions behave when interacting with the solid. Therefore, we assert that $\hat{O}$ corresponds to some observable for the mobile ions so that, in the particle-mode space, it becomes $\hat{O}\rightarrow \hat{O}\otimes \hat{1}$.

To proceed from Eq.~\eqref{eqn:O_tau_general}, we employ the standard path integral approach of rewriting the time evolution operators as $e^{\pm\frac{i\hat{H}_i\tau}{\hbar}}=\left(e^{\pm\frac{i \hat{H}_i \Delta}{\hbar}}\right)^{N-1}$ for $\tau / (N - 1) = \Delta$ and $N\rightarrow \infty$, and inserting identity operators between the multiples:
\begin{align}
    \langle\hat{O}\rangle (\tau) 
    &= 
    \frac{1}{\mathrm{Tr}\left[\hat{\rho}_0\right]} 
    \sum_{n,\mathbf{r}} 
    \langle \mathbf{r},n|\mathbf{1}_-^1 e^{\frac{i \hat{H}_i \Delta}{\hbar}}
    \mathbf{1}_-^2
    \dots
    e^{\frac{i \hat{H}_i \Delta}{\hbar}}
    \mathbf{1}_-^N
    \nonumber
    \\
    &\times 
    \hat{O}
    \mathbf{1}_+^N
    e^{-\frac{i \hat{H}_i \Delta}{\hbar}}
    \dots
    e^{-\frac{i \hat{H}_i \Delta}{\hbar}}
    \mathbf{1}_+^1
    \hat{\rho}_0
    |\mathbf{r},n\rangle
    \,.
    \label{eqn:O_tau_Trotter}
\end{align}
The subscript on the identity operators indicates whether they are located on the right ($+$) or on the left ($-$) of $\hat{O}$. The superscript indicates how many time steps from $t = 0$ the operator is positioned (note that the negative sign in the exponential moves the time forward, and the positive one moves it backward).

It is convenient to choose the identity operators composed of bosonic coherent states:
\begin{equation}
    \mathbf{1}_\pm^j\equiv \int d\mathbf{r}_\pm^j \int \frac{d\bar{\mathbf{s}}_{\pm}^j \, d\mathbf{s}_{\pm}^j}{\pi} e^{-\bar{\mathbf{s}}_\pm^j\mathbf{s}_{\pm}^j}|\mathbf{r}_\pm^j,\mathbf{s}_\pm^j \rangle \langle \mathbf{r}_\pm^j,\mathbf{s}_\pm^j|\,.
    \label{eqn:Identity}
\end{equation}
Here, $\mathbf{s}^j_\pm$ is a column vector of complex numbers, one for each vibrational mode $s$ and $\bar{\mathbf{s}}_\pm^j$ is its conjugate transpose. Because $|\mathbf{r}_\pm^j,\mathbf{s}_\pm^j \rangle = |\mathbf{r}_\pm^j\rangle\otimes |\mathbf{s}_\pm^j \rangle$, Eq.~\eqref{eqn:Identity} can be regarded as a direct product of two identities.

In the first part of the right-hand side of Eq.~\eqref{eqn:O_tau}, we have a term $\langle\mathbf{r},n| \mathbf{r}^{1}_-,\mathbf{s}^1_-\rangle \langle \mathbf{r}^{1}_-,\mathbf{s}^1_-|\dots$. Because $\langle\mathbf{r},n| \mathbf{r}^{1}_-,\mathbf{s}^1_-\rangle$ is a number, we can move it to the right of Eq.~\eqref{eqn:O_tau} to obtain $\sum_{\mathbf{r},n} \dots\hat{\rho}_0|\mathbf{r},n\rangle\langle\mathbf{r},n| \mathbf{r}^{1}_-,\mathbf{s}^1_-\rangle$. This step allows us to eliminate the summation over $\mathbf{r}$ and $n$ because it has the form of a resolution identity. With this rearrangement, we obtain
\begin{align}
    &\langle\hat{O}\rangle (\tau) 
    = 
    \int \frac{\mathcal{D}\left(\dots\right) }{\mathrm{Tr}\left[\hat{\rho}_0\right]}
    \exp\left[-\sum_{j = 1}^N \left(\bar{\mathbf{s}}_-^j\mathbf{s}_-^j + \bar{\mathbf{s}}_+^j\mathbf{s}_+^j\right)\right]
    \nonumber
    \\
    \times &
    \langle \mathbf{r}_-^1,\mathbf{s}_-^1| 
    e^{\frac{i \hat{H}_i \Delta}{\hbar}}
    | \mathbf{r}_-^2,\mathbf{s}_-^2\rangle\langle \mathbf{r}_-^2,\mathbf{s}_-^2|
    \dots
    e^{\frac{i \hat{H}_i \Delta}{\hbar}}
    | \mathbf{r}_-^N,\mathbf{s}_-^N\rangle
    \nonumber
    \\
    \times & \langle \mathbf{r}_+^N,\mathbf{s}_+^N|
    e^{-\frac{i \hat{H}_i \Delta}{\hbar}}
    \dots
    | \mathbf{r}_+^2,\mathbf{s}_+^2\rangle\langle \mathbf{r}_+^2,\mathbf{s}_+^2|
    e^{-\frac{i \hat{H}_i \Delta}\hbar}
    | \mathbf{r}_+^1,\mathbf{s}_+^1\rangle
     \nonumber
     \\
     \times&
    \langle \mathbf{r}_-^N,\mathbf{s}_-^N|
    \hat{O}
    | \mathbf{r}_+^N,\mathbf{s}_+^N\rangle\langle \mathbf{r}_+^1,\mathbf{s}_+^1|
    \hat{\rho}_0
    |\mathbf{r}_-^1,\mathbf{s}_-^1\rangle
    \,,
    \label{eqn:O_tau}
\end{align}
where $\mathcal{D}\left(\dots\right)$ contains all the differentials and $\pi^{-1}$ prefactors of the integrals.

Computing the matrix elements, as shown in Appendix~\ref{sec:Matrix_Elements}, followed by the integration over the mode fields, as outlined in Appendix~\ref{sec:Field_Integration}, gives
\begin{widetext}
\begin{align}
    &\langle \hat{O}\rangle (\tau) 
     =
    \frac{1}{\mathrm{Tr}\left[\hat{\rho}_m\right]}
    \int \mathcal{D}\left(\dots\right)
    \langle \mathbf{r}_-^N|\hat{O}|\mathbf{r}_+^N\rangle \langle \mathbf{r}^1_+|\hat{\rho}_m|\mathbf{r}^{1}_- \rangle
    \left|\frac{\mathbf{M}}{2\pi  \Delta \hbar}\right|^{N-1}
    \prod_s \exp\left[-\frac{1}{2}
    \coth\left(\frac{\beta\hbar\Omega_s}{2}\right)
    Q_sQ^*_s
    \right]
    \nonumber
    \\
    \times&
    \prod_s \exp\Bigg[i\frac{\Delta^2}{\hbar^2}\sum_{ln = 1}^N 
    \sin\left(\Delta\Omega_s(n - l)\right)
    \left(Y_s(\mathbf{r}_+^n) - Y_s(\mathbf{r}_-^n)\right)
    \Theta(n - l)
    \left( Y_s(\mathbf{r}_+^l) + Y_s(\mathbf{r}_-^l) + 2W_s(l\Delta)\right)
    \Bigg]
    \nonumber
    \\
    \times &
    \prod_{j = 1}^{N - 1}
    \exp\Bigg[\sum_{\sigma = \pm}\sigma\frac{i\left(\mathbf{r}_\sigma^{j+1} - \mathbf{r}_\sigma^j\right)^T\mathbf{M}\left(\mathbf{r}_\sigma^{j+1} - \mathbf{r}_\sigma^j\right)}{2\Delta\hbar}-\sigma\frac{i\Delta}{\hbar} U(\mathbf{r}_\sigma^j)
    \Bigg]
    \label{eqn:O_tau_integrated}
\end{align}
\end{widetext}
with
\begin{equation}
     Q_s 
    = \frac{\Delta}{\hbar}\sum_{l = 1}^N 
    e^{-i\Delta \Omega_s l}
    \left[
    Y_s(\mathbf{r}_+^l) - Y_s(\mathbf{r}_-^l)
    \right]
    \,.
    \label{eqn:Q}
\end{equation}
Note that the $\hat{\rho}$ in the trace is that of the mobile particles since the mode portion is cancelled by the field integration. 

\subsection{Hubbard-Stratonovich Transformation}
\label{sec:HS}

As the next step, we employ the Hubbard-Stratonovich transformation to separate the product $Q_sQ_s^*$ in the exponential. The first step is to define a unity
\begin{align}
    1&\equiv \frac{2}{\pi}
     \tanh\left(\frac{\beta\hbar\Omega_s}{2}\right)
     \nonumber
     \\
     &\times \int d\xi_sd\xi^*_s  \exp\left[-2\xi^*_s
     \tanh\left(\frac{\beta\hbar\Omega_s}{2}\right)\xi_s\right]
     \label{eqn:Unity}\,,
\end{align}
where $\xi_s$ is a complex variable and $\xi_s^*$ is its conjugate so the integration takes place over the entire complex plane. The equality holds even if one shifts $\xi_s$ and $\xi_s^*$ by arbitrary independent complex numbers. Hence, we choose $\xi_s \rightarrow \xi_s+\frac{i}{2} \coth\left(\frac{\beta\hbar\Omega_s}{2}\right) Q_s$ and $\xi_s^* \rightarrow \xi_s^*+\frac{i}{2} \coth\left(\frac{\beta\hbar\Omega_s}{2}\right) Q_s^*$. Inserting this unity into Eq.~\eqref{eqn:O_tau_integrated} and rearranging the terms yields
\begin{widetext}
\begin{align}
    &\langle \hat{O}\rangle (\tau) 
    =
    \int \mathcal{D}\left(\dots\right) \langle \mathbf{r}_-^N|\hat{O}|\mathbf{r}_+^N\rangle 
     \frac{\langle \mathbf{r}^1_+|\hat{\rho}_m|\mathbf{r}^{1}_-\rangle}{\mathrm{Tr}\left[\hat{\rho}_m\right]}
    \left|\frac{\mathbf{M}}{2\pi  \Delta \hbar}\right|^{N-1}
    \prod_s
    \frac{2}{\pi}
     \tanh\left(\frac{\beta\hbar\Omega_s}{2}\right)\int d\xi_sd\xi^*_s
     \exp\left[-
     2
     \xi^*_s
     \tanh\left(\frac{\beta\hbar\Omega_s}{2} \right)
     \xi_s
     \right]
    \nonumber
    \\
    \times&
    \prod_s 
   \exp\Bigg\{-i\frac{\Delta}{\hbar}\sum_{n = 1}^N 
   \left[
    Y_s(\mathbf{r}_+^n) - Y_s(\mathbf{r}_-^n)
    \right]
    \left[\xi_s e^{i\Delta \Omega_s n}
    +\xi^*_s e^{-i\Delta \Omega_s n }\right]
    \Bigg\}
    \nonumber
    \\
    \times&
    \prod_s \exp\Bigg[i\frac{\Delta^2}{\hbar^2}\sum_{ln = 1}^N 
    \sin\left(\Delta\Omega_s(n - l)\right)
    \left(Y_s(\mathbf{r}_+^n) - Y_s(\mathbf{r}_-^n)\right)
    \Theta(n - l)
    \left( Y_s(\mathbf{r}_+^l) + Y_s(\mathbf{r}_-^l) + 2W_s(l\Delta)\right)
    \Bigg]
    \nonumber
     \\
     \times &
   \prod_{j = 1}^{N - 1}
    \exp\Bigg[\sum_{\sigma = \pm}\sigma\frac{i\left(\mathbf{r}_\sigma^{j+1} - \mathbf{r}_\sigma^j\right)^T\mathbf{M}\left(\mathbf{r}_\sigma^{j+1} - \mathbf{r}_\sigma^j\right)}{2\Delta\hbar}
    -\sigma\frac{i\Delta}{\hbar} U(\mathbf{r}_\sigma^j)
    \Bigg] 
    \,.
    \label{eqn:O_tau_HS}
\end{align}
\end{widetext}

At this point, it might be unclear why we used the decoupling. After all, it appears to have reinserted a phononic fields that we just integrated out. In fact, it is not quite that: this decoupling eliminated a specific type of term (the product of differences of $Y_s$), the benefit of which will become apparent when we treat the system semiclassically.

\subsection{Semiclassical Approximation}
\label{sec:Semi_Classical}

We start by rewriting the coordinates as $\mathbf{r}_\pm^j = (\mathbf{r}_c^j \pm \mathbf{r}_q^j)/\sqrt{2}$. Expanding the terms in the exponential in Eq.~\eqref{eqn:O_tau_HS} to the leading order in $\mathbf{r}_q^j$ gives:
\begin{align}
    & U(\mathbf{r}_+^j) - U(\mathbf{r}_-^j) 
     \rightarrow
     \sqrt{2}\left[\nabla U(\mathbf{r}_c^j/\sqrt{2})\right]^T \mathbf{r}_q^j\,,
     \nonumber
    \\
   & Y_s(\mathbf{r}_+^j) +  Y_s(\mathbf{r}_-^j)
    \rightarrow
    2Y_s(\mathbf{r}_c^j/\sqrt{2})\,,
    \nonumber
    \\
   & Y_s(\mathbf{r}_+^j) -  Y_s(\mathbf{r}_-^j)
    \rightarrow
     \sqrt{2} 
     \left[\nabla Y_s(\mathbf{r}_c^j/\sqrt{2})\right]^T\mathbf{r}_q^j
     \,,
     \nonumber
     \\
     &\frac{1}{2}\sum_{\sigma=\pm} \sigma\left(\mathbf{r}_\sigma^{j+1} - \mathbf{r}_\sigma^j\right)^T
    \mathbf{M}
    \left(\mathbf{r}_\sigma^{j+1} - \mathbf{r}_\sigma^j\right)
    \nonumber
    \\
    \rightarrow &
    \left(\mathbf{r}_q^{j+1}-\mathbf{r}_q^{j}\right)^T
    \mathbf{M}
     \left(\mathbf{r}_c^{j+1}-\mathbf{r}_c^{j}\right)
     \,.
     \label{eqn:rq_expansion}
\end{align}

Following this expansion, we write
\begin{align}
    &\prod_{j = 1}^{N - 1}
    \exp\left[\sum_{\sigma = \pm}\sigma\frac{i\left(\mathbf{r}_\sigma^{j+1} - \mathbf{r}_\sigma^j\right)^T\mathbf{M}\left(\mathbf{r}_\sigma^{j+1} - \mathbf{r}_\sigma^j\right)}{2\Delta\hbar}
    \right] 
    \nonumber
    \\
    \rightarrow&
    \exp\left[
    \sum_{j = 2}^{N-1}i\frac{\mathbf{r}_q^{j}\cdot\mathbf{M}\left(2\mathbf{r}_c^{j}-\mathbf{r}_c^{j-1}-\mathbf{r}_c^{j+1}\right)}{\Delta\hbar}
    \right]
\end{align}
in Eq.~\eqref{eqn:O_tau_HS}, where we set $\mathbf{r}_q$ to vanish at the endpoints of the time contour. Inserting the rest of expressions from Eq.~\eqref{eqn:rq_expansion} into Eq.~\eqref{eqn:O_tau_HS} and integrating over $\mathbf{r}_q$ results in

\begin{widetext}
\begin{align}
    \langle \hat{O}\rangle (\tau) 
    &=
    \frac{1}{\mathrm{Tr}\left[\hat{\rho}_m\right]}
    \left|\frac{\mathbf{M}}{2\pi  \Delta \hbar}\right|^{N-1}
    \prod_s
     \frac{2}{\pi}
     \tanh\left(\frac{\beta\hbar\Omega_s}{2}\right)
     \int d\xi_sd\xi^*_s
     \exp\left[-
     2
     \xi^*_s
     \tanh\left(\frac{\beta\hbar\Omega_s}{2} \right)
     \xi_s
     \right]
    \nonumber
    \\
    &\times
    \int\prod_j d\mathbf{r}_c^j
    \prod_{n=2}^{N-2}
    \delta\Bigg\{
    \frac{
     \mathbf{M}\left(2\mathbf{r}_c^{j}-\mathbf{r}_c^{j-1}-\mathbf{r}_c^{j+1}\right)
    }{\hbar\Delta}-\frac{\Delta}{\hbar}\sqrt{2}\nabla U(\mathbf{r}_c^j/\sqrt{2})
    -\frac{\Delta}{\hbar}
    \sqrt{2}\nabla Y_s(\mathbf{r}_c^n/\sqrt{2})
    \left[\xi_s e^{i\Delta \Omega_s n}
    +\xi^*_s e^{-i\Delta \Omega_s n }\right]
    \nonumber
    \\
    &+
    \frac{\Delta^2}{\hbar^2}\sum_{l = 1}^N 
    \sin\left(\Delta\Omega_s(n - l)\right)
     \sqrt{2}\nabla Y_s(\mathbf{r}_c^n/\sqrt{2})
    \Theta(n - l)
    2\left[Y_s(\mathbf{r}_c^l/\sqrt{2})+W_s(l\Delta)\right]
    \Bigg\}
    \langle \mathbf{r}_c^N|\hat{O}|\mathbf{r}_c^N\rangle \langle \mathbf{r}^1_c|\hat{\rho}_p|\mathbf{r}^{1}_c\rangle 
    \,.
    \label{eqn:O_tau_SemiClassical}
\end{align}
\end{widetext}

Relabeling $\frac{\mathbf{r}_c}{\sqrt{2}}\rightarrow \mathbf{r}$ in Eq.~\eqref{eqn:O_tau_SemiClassical}, one can identify the equation of motion for the ions inside the Dirac delta function
\begin{align}
    &\mathbf{M}\frac{\mathbf{r}_{n-1}-2\mathbf{r}_{n}+\mathbf{r}_{n+1}}{\Delta^2}
    =-\nabla U(\mathbf{r}_n,n\Delta)
     \nonumber
     \\
     &+ 
    2\frac{\Delta}{\hbar}\sum_s 
  \nabla Y_s(\mathbf{r}_n)
    \sum_{l = 1}^n
    \sin\left[\Delta \Omega_s \left(n-l\right)\right]
     \left[ Y_s(\mathbf{r}_l)+W_s(l\Delta)\right]
    \nonumber
     \\
     &\underbrace{- \sum_s 
  \nabla Y_s(\mathbf{r}_n)
  \left[
   e^{i\Delta \Omega_s n}
   \xi_s
     +
   e^{-i\Delta \Omega_s n }
    \xi^*_s\right]}_{\tilde{\mathbf{f}}_n}\,.
    \label{eqn:EOM}
\end{align}

The first line of Eq.~\eqref{eqn:EOM} describes the motion of the ions in a $t$- and $\mathbf{r}$-dependent potential. The second line introduces the recoil: mobile ions experience a force at time $n$ as a consequence of the stationary ions being perturbed by the external potential and the mobile ions at time $l$. Finally, the third line describes the stochastic thermal force with the probability distribution of $\zeta_s$ given by the integrand of Eq.~\eqref{eqn:Unity}. The right-hand side of Eq.~\eqref{eqn:EOM} expression agrees with Eq.~\eqref{eqn:Classical_Force} obtained using the classical approach, but Eq.~\eqref{eqn:EOM} also explicitly gives the temperature dependence of the homogeneous solution of the framework's equations of motion.

\section{Drift and Diffusion}
\label{sec:Drift_Diffusion}

\subsection{Fluctuation-dissipation in solids}
\label{sec:FDT_solids}

Because the thermal force originates from the vibrations of the solid, it exhibits a finite correlation in time, which can be quantified using the correlation tensor

\begin{align}
    \langle \tilde{\mathbf{f}}_n\otimes\tilde{\mathbf{f}}_l\rangle
    =
    \sum_s 
    &
     \nabla Y_s\left(\mathbf{r}_n\right)
     \otimes
     \nabla Y_s\left(\mathbf{r}_l\right)
     \nonumber
     \\
     &\times
     \coth\left(\frac{\beta\hbar\Omega_s}{2}\right)\cos\left[\Delta \Omega_s\left(n - l\right)\right]\,,
     \label{eqn:Corr_Tensor}
\end{align}
see Appendix~\ref{sec:Correlation_Tensor} for the derivation.

In accordance with the fluctuation-dissipation theorem, the recoil and thermal noise terms in Eq.~\eqref{eqn:EOM} form a fluctuation-dissipation pair as they originate from the same physical phenomenon, namely the interaction of mobile ions with the framework. To bring this relationship to a more familiar form, let us consider a scenario where the there is no external perturbation, eliminating the $W_s$ term and the time-dependence of $U$ from Eq.~\eqref{eqn:EOM}. As shown in Appendix~\ref{sec:Recoil_Term}, for $\Delta\rightarrow 0$, we can write the recoil term as

\begin{align}
    &2\frac{\Delta}{\hbar}\sum_s 
  \nabla Y_s(\mathbf{r}_n)\sum_{l = 1}^n
     \sin\left[\Delta\Omega_s\left(n  - l\right)\right] Y_s(\mathbf{r}_l)
     \nonumber
     \\
     \approx&
    \nabla \sum_s \frac{Y_s^2(\mathbf{r}_n)}{\hbar\Omega_s} 
    -
   2\sum_s 
  \nabla Y_s(\mathbf{r}_n) \frac{\cos\left[\Delta\Omega_s n\right]}{\hbar\Omega_s} Y_s(\mathbf{r}_{1})
    \nonumber
    \\
    -& 
   2\frac{\Delta}{\hbar}\sum_s 
   \sum_{l = 1}^{n-1} \frac{\cos\left[\Delta\Omega_s\left(n  - l\right)\right]}{\Omega_s} 
   \nabla Y_s(\mathbf{r}_n)\otimes \nabla Y_s(\mathbf{r}_{l})
     \dot{\mathbf{r}}_{l}
     \label{eqn:Recoil}
   \,.
\end{align}
The first term describes the softening of the potential $U$ due to the solid's elasticity, as one can see by combining it with the first term on the r.h.s. of Eq.~\eqref{eqn:EOM}. The second one is the boundary term carrying the information about the initial configuration and vanishing as $n\rightarrow \infty$.

Comparing the final term to Eq.~\eqref{eqn:Corr_Tensor} for $T\gg 1$ with $\coth(\beta\hbar\Omega_s / 2) \approx 2\beta^{-1}/\hbar\Omega$, shows that it can be written as $-\beta \Delta\sum_{l = 1}^{n-1} \langle \tilde{\mathbf{f}}_n\otimes\tilde{\mathbf{f}}_l\rangle \dot{\mathbf{r}}_{l}$. This relation between the recoil term and the noise correlation tensor is a consequence of the fluctuation-dissipation theorem. To make this connection more explicit, we write the high-$T$ version of the correlation tensor explicitly as
\begin{align}
    \langle \tilde{\mathbf{f}}_n\otimes\tilde{\mathbf{f}}_l\rangle
    =
    \sum_s 
    &
     \nabla 
    \left[\nabla_{\mathbf{u}^0} U\left(\hat{\mathbf{r}},\mathbf{u}^0\right)
    \right]^T
    \mathbf{m}^{-1/2}
    \boldsymbol{\varepsilon}_{s}
    \nonumber
    \\
     \otimes
     &
     \nabla 
    \left[\nabla_{\mathbf{u}^0} U\left(\hat{\mathbf{r}},\mathbf{u}^0\right)
    \right]^T
    \mathbf{m}^{-1/2}
    \boldsymbol{\varepsilon}_{s}
     \nonumber
     \\
     \times&
     \frac{\cos\left[\Delta \Omega_s\left(n - l\right)\right]}{\beta \Omega_s^2}\,.
     \label{eqn:Corr_Tensor_high_T}
\end{align}
If the system is three-dimensional, the vibrational modes at low energy have a density of states that is quadratic in $\Omega_s$, corresponding to acoustic modes. This density of states cancels the $\Omega_s^2$ term in the denominator, preventing a low-energy divergence seen in one- and two-dimensional systems. This cancellation means that the oscillatory cosine term strongly suppresses the correlation tensor for $n\neq l$. In the case of low-dimensional systems, one needs to suppress the divergence by, for example, confining the system in an external potential, eliminating the zero-frequency modes.

If the the velocities and the positions of the mobile ions change on much longer time scales than the decay of the correlation tensor, we can replace $\dot{\mathbf{r}}_l\rightarrow \dot{\mathbf{r}}_n$ and $\mathbf{r}_l\rightarrow \mathbf{r}_n$, and, for $n\gg 1$, extend the lower bound of the $l$ summation to $-\infty$ in Eq.~\eqref{eqn:Recoil}. We can then write the last term of Eq.~\eqref{eqn:Recoil} as $-\boldsymbol{\gamma}_n\dot{\mathbf{r}}_{n}$ with

\begin{align}
    2k_BT\boldsymbol{\gamma}_n =  2\Delta\sum_{l =- \infty}^{n} 
    \langle \tilde{\mathbf{f}}_n\otimes\tilde{\mathbf{f}}_l\rangle =  \Delta\sum_{l = -\infty}^{\infty} 
    \langle \tilde{\mathbf{f}}_n\otimes\tilde{\mathbf{f}}_l\rangle \,.
    \label{eqn:FDT}
\end{align}
The second equality holds because, after the $\mathbf{r}_l\rightarrow \mathbf{r}_n$ replacement in $\langle \tilde{\mathbf{f}}_n\otimes\tilde{\mathbf{f}}_l\rangle$, $l$ appears only in the cosine term.

One can identify $\boldsymbol{\gamma}_n$ as the position-dependent drag tensor. The relation between $\boldsymbol{\gamma}_n$ and the noise correlation tensor in Eq.~\eqref{eqn:FDT} is precisely the form required by the fluctuation-dissipation theorem in the Langevin limit, where rapid light particles of the medium impact slow impurities in a white-noise-like manner. Performing the summation over $l$ yields
\begin{equation}
    \boldsymbol{\gamma}_n
      =2\pi
     \sum_s 
     \nabla Y_s\left(\mathbf{r}_n\right)
     \otimes
     \nabla Y_s\left(\mathbf{r}_n\right)
     \frac{\delta(\Omega_s)}{\hbar\Omega_s}\,,
     \label{eqn:gamma}
\end{equation}
demonstrating that, in the Langevin regime, mobile particles dissipate energy via low-frequency framework modes.

Because, in the limit of $\Omega_s \rightarrow 0$, all crystal atoms move with the same phase and amplitude, leading to $\mathbf{m}^{-1/2}\boldsymbol{\varepsilon}_{s}\rightarrow \mathbf{1}_{I}\otimes \boldsymbol{\varepsilon}_{s}/(\sqrt{mL})$, where $m$ is the mass of all the atoms in the unit cell, $L$ is the number of unit cells in the system, and the newly-defined $\boldsymbol{\varepsilon}_s$ is a $D$-dimensional polarization vector. This form results in a substantial simplification:

\begin{align}
        Y_{s}(\mathbf{r}) &=
    \sqrt{\frac{\hbar}{2\Omega_s}}
    \left[\nabla_{\mathbf{u}^0} U\left(\mathbf{r},\mathbf{u}^0\right)
    \right]^T\left[\mathbf{1}_{I}\otimes \boldsymbol{\varepsilon}_{s}/(\sqrt{mL})\right]
    \nonumber
    \\
    &=
    \sqrt{\frac{\hbar}{2\Omega_s}}
    \left[\left(\sum_l\nabla_{\mathbf{u}^0_l}\right) U\left(\mathbf{r},\mathbf{u}^0\right)
    \right]^T\left[ \boldsymbol{\varepsilon}_{s}/(\sqrt{mL})\right]\,.
\end{align}
One can identify the term in the parentheses as the gradient with respect to the uniform shift of all the stationary ions. 

\subsection{Superionic Conduction}
\label{eqn:Superionic_Conduction}

It is convenient to study the motion of the mobile ions in superionic conductors using the independent-ion approximation by setting $\mathbf{r}$ to represent the location of a single mobile ion. In the presence of an external constant force $\mathbf{F}$, the equation of motion of a single ion becomes $M\ddot{\mathbf{r}} = -\nabla \bar{U}(\mathbf{r}) + \mathbf{F} - \boldsymbol{\gamma}(\mathbf{r}){\color{black}{\dot{\mathbf{r}}}} + \tilde{\mathbf{f}}$. The bar over $U$ indicates that this potential includes the softening effects of the first term in Eq.~\eqref{eqn:Recoil}. In this approximation, a uniform shift of the stationary ions is equivalent to a shift of the mobile ion in the opposite direction, yielding

\begin{align}
    \nabla Y_s(\mathbf{r})
    =
     - \sqrt{\frac{\hbar}{2mL\Omega_s}}
    \left[\mathbf{H}_{\mathbf{r}} U\left(\mathbf{r}\right)
    \right]\boldsymbol{\varepsilon}_{s}\,,
\end{align}
where $\mathbf{H}_\mathbf{r}$ is the Hessian operator. Explicitly, the drag matrix is given by

\begin{equation}
    \boldsymbol{\gamma}
      =
     \left[\mathbf{H}_{\mathbf{r}} U\left(\mathbf{r}\right)
    \right]
    \left[
    \sum_s 
     \frac{2\pi}{2mL}
     \frac{\delta(\Omega_s)}{\Omega_s^2}
    \boldsymbol{\varepsilon}_{s}\otimes \boldsymbol{\varepsilon}_{s}\right]
    \left[\mathbf{H}_{\mathbf{r}} U\left(\mathbf{r}\right)
    \right]\,.
\end{equation}

For the expression in the brackets, we write

\begin{align}
    & \frac{2\pi}{2m}
     \sum_s 
     \delta(\Omega_s)
    \frac{\boldsymbol{\varepsilon}_{s}
     \otimes\boldsymbol{\varepsilon}_{s}}{\Omega_s^2L}
     =
     \frac{2\pi}{2m}
     \sum_{b,\mathbf{q}} 
     \delta(\Omega_{b,\mathbf{q}})
    \frac{\boldsymbol{\varepsilon}_{b,\mathbf{q}}
     \otimes\boldsymbol{\varepsilon}_{b,\mathbf{q}}}{\Omega_{b,\mathbf{q}}^2L}
     \nonumber
     \\
     =&
     \frac{2\pi V }{2(2\pi)^3m L}
     \sum_{b}\int d\mathbf{q} 
    \frac{\boldsymbol{\varepsilon}_{b,\mathbf{q}}
     \otimes\boldsymbol{\varepsilon}_{b,\mathbf{q}}}{\Omega_{b,\mathbf{q}}^2}
     \delta(\Omega_{b,\mathbf{q}})\,.
\end{align}
Here, $b$ labels the phonon branch and $V$ is the volume of the system. Because $m L$ is the total mass of the system, $V / (mL)$ gives the density $\rho$. In the $\mathbf{q}\rightarrow 0$ limit, $\Omega_{b,\mathbf{q}}\rightarrow v_b(\theta,\phi)q$, where $v_b(\theta,\phi)$ is the direction-dependent sound velocity for branch $b$.

If we make $v_b$ isotropic (as one can expect it to be in a polycrystalline macroscopic sample), the integral can be written as

\begin{widetext}
\begin{align}
   & \frac{1}{2(2\pi)^2\rho}
     \sum_{b}\int d\mathbf{q} 
    \frac{\boldsymbol{\varepsilon}_{b,\mathbf{q}}
     \otimes\boldsymbol{\varepsilon}_{b,\mathbf{q}}}{v_b^2 q^2}
     \delta(v_b q)
     =
     \frac{1}{2(2\pi)^2\rho}
     \sum_{b}\oint d\phi \int d\theta \sin\theta \int dq\,q^2
    \frac{\boldsymbol{\varepsilon}_{b,\mathbf{q}}
     \otimes\boldsymbol{\varepsilon}_{b,\mathbf{q}}}{v_b^2 q^2}
     \delta(v_b q)
     \nonumber
     \\
     =&
     \frac{1}{2(2\pi)^2\rho}
     \left[
     \int d\boldsymbol{\Omega}\,
     \hat{\mathbf{r}}\otimes\hat{\mathbf{r}}
     \int dq
    \frac{\delta(v_L q)
     }{v_L^2 }
     +
     \int d\boldsymbol{\Omega}\,
     \left(1 - \hat{\mathbf{r}} \otimes\hat{\mathbf{r}}\right)
     \int dq
    \frac{\delta(v_T q)
     }{v_T^2 }
     \right]
     \nonumber
     \\
     =&
     \frac{1}{2(2\pi)^2\rho}\frac{4\pi}{3}
     \left(
    \frac{1}{2v_L^3 }
     +
    \frac{1}{v_T^3 }
     \right)
     \nonumber
     \\
      =&
     \frac{1}{12\pi\rho}
     \left(
    \frac{1}{v_L^3 }
     +
    \frac{2}{v_T^3 }
     \right)\,,
\end{align}
\end{widetext}
where $d\boldsymbol{\Omega}$ denotes the integration over the solid angle, while $v_L$ and $v_T$ are the speeds of sound for longitudinal and transverse modes, respectively. At each momentum $\mathbf{q}$, there are three phonon branches: a longitudinal one propagating in the $\hat{\mathbf{r}}$ direction and two transverse ones propagating in $\hat{\boldsymbol{\phi}}$ and $\hat{\boldsymbol{\theta}}$ directions. It is these branches that give rise to the three terms in the second line above.

Combining the results yields

\begin{align}
    \boldsymbol{\gamma}
   =
    \frac{1}{12\pi\rho}
     \left(
    \frac{1}{v_L^3 }
     +
    \frac{2}{v_T^3 }
     \right)\left[ \mathbf{H}_\mathbf{r}U\left(\mathbf{r}\right)\right]^2\,.
     \label{eqn:gamma_Hessian}
\end{align}

The expression for $\boldsymbol{\gamma}$ suggests what types of systems would lead to the smallest energy dissipation of mobile ions. One can see that dense (high $\rho$) and stiff (high $v_L$ and $v_T$) materials yield a lower $\boldsymbol{\gamma}$. Furthermore, the movement of the ion is dissipationless in regions of the potential where the Hessian vanishes, which are the saddle point regions of the periodic potential in the unit cell of the crystal.

In the long-time limit, the inertia term $M\ddot{\mathbf{r}}$ and the random force $\tilde{\mathbf{f}}$ can be dropped from the single-ion equation of motion, leading to $\boldsymbol{\gamma}(\mathbf{r}){\color{black}\dot{\mathbf{r}}}= -\nabla \bar{U}(\mathbf{r})+ \mathbf{F}$ or, alternatively, ${\color{black}\dot{\mathbf{r}}}= \boldsymbol{\gamma}^{-1}\left( \mathbf{r}\right) \left[-\nabla \bar{U}(\mathbf{r})+\mathbf{F} \right]$. {\color{black}As the mobile ion moves in response to the applied force $\mathbf{F}$, it will speed up and slow down periodically due to the spatially varying potential landscape.} Because the potential and the drag terms have the lattice periodicity, it is reasonable to expect that ${\color{black}\dot{\mathbf{r}}}$ will exhibit the same variation. {\color{black}Hence, we define a velocity field $\mathbf{v}(\mathbf{r})$, written in Fourier space as}

\begin{align}
   \mathbf{v}_\mathbf{K} &= \left(2\pi\right)^{3/2}\sum_{\mathbf{K}'}
   \boldsymbol{\gamma}^{-1}_{\mathbf{K} - \mathbf{K}'} \left[-i\mathbf{K}' \bar{U}_{\mathbf{K}'}+\mathbf{F} \delta_{0,\mathbf{K}'} \right]\,.
    \label{eqn:Drift}
\end{align}
Setting $\mathbf{K}\rightarrow 0$ yields the drift velocity $\mathbf{v}_0$. Because $\bar{U}_\mathbf{K} = \bar{U}_{-\mathbf{K}}$, $\boldsymbol{\gamma}_\mathbf{K} = \boldsymbol{\gamma}_{-\mathbf{K}}$, and $\left(2\pi\right)^{3/2} \boldsymbol{\gamma}^{-1}_{\mathbf{K}=0} = \langle \boldsymbol{\gamma}^{-1}\rangle$ is the average of the inverse drag tensor,

\begin{equation}
    \mathbf{v}_\mathrm{drift} = \mathbf{v}_{\mathbf{K}\rightarrow 0}=\langle \gamma^{-1}\rangle q\mathbf{E}\,.
    \label{eqn:v_drift}
\end{equation}
From the above expression we can readily obtain the ion mobility:

\begin{equation}
    \mu = \frac{ \mathbf{v}_\mathrm{drift}}{E} = q\langle \gamma^{-1}\rangle\,,
    \label{eqn:mu}
\end{equation}
where $\langle \gamma^{-1}\rangle$ can be computed from first principles for any crystal lattice. 

Note that this is the mobility per mobile ion --- similar to the definition of the mobility of electrons and holes in metals or semiconductors, which naturally excludes the electrons that are not involved in the transport. This mobility originates from the ion-lattice collision frequency, which depends on the the structure of the medium, but not on the temperature, reminiscent of the Drude model for electrons. 

{\color{black}
It might appear counter-intuitive that temperature does not appear in $\gamma$ given that the lattice vibrations depend on the temperature. Moreover, experimentally measured conductivity does indeed exhibit temperature dependence. To resolve this apparent contradiction, we reiterate that $\gamma$ is related to drag, which is a purely dissipative process, caused by ionic collisions. The (random) thermal motion of the lattice does not, on average, change the amount of energy transferred from the mobile ions to the crystal, explaining the lack of $T$ in $\gamma$. On the other hand, increasing the temperature can introduce more mobile ions to the system as larger thermal vibrations liberate more of them from the local energy minima. Thus, the experimentally observed temperature dependence of the conductivity stems not from increased mobility, but rather from increased ionic density.
}

Knowing the potential landscape $U({\bf r})$, which can be obtained for example from density functional theory calculations, it is possible to obtain estimates for the ionic mobilities from Eq.~\eqref{eqn:gamma_Hessian}.
One of the advantages of this method is that, by construction, the whole potential surface is taken into account. In contrast, in molecular dynamics a satisfactory sampling of the configuration space requires, in practice, very long integration times. In theory, the expectation that in molecular dynamics the system will eventually pass through all possible states, if allowed to evolve indefinitely, is based on the ergodic hypothesis, which states that the time average equals the ensemble average\cite{coveney2016calculation}.
Such an approach is not necessarily valid for non-equilibrium systems.

Additionally, molecular dynamics simulations require the choice of a time integration step that is small enough to guarantee the convergence of the integrated coordinates. Thus, it is impractical to simulate using the same method diffusion or drag in conditions where the conductivity varies by orders of magnitude.
In contrast, the present approach measures energies, rather than time, and is therefore widely applicable to different materials.

\section{Numerical Results}
\label{sec:Numerical_Results}

In order to substantiate our results, we make use of \textit{ab initio} density functional theory (DFT) to calculate the ion mobility for some crystalline electrolytes.

\begin{figure}
    \centering
    \includegraphics{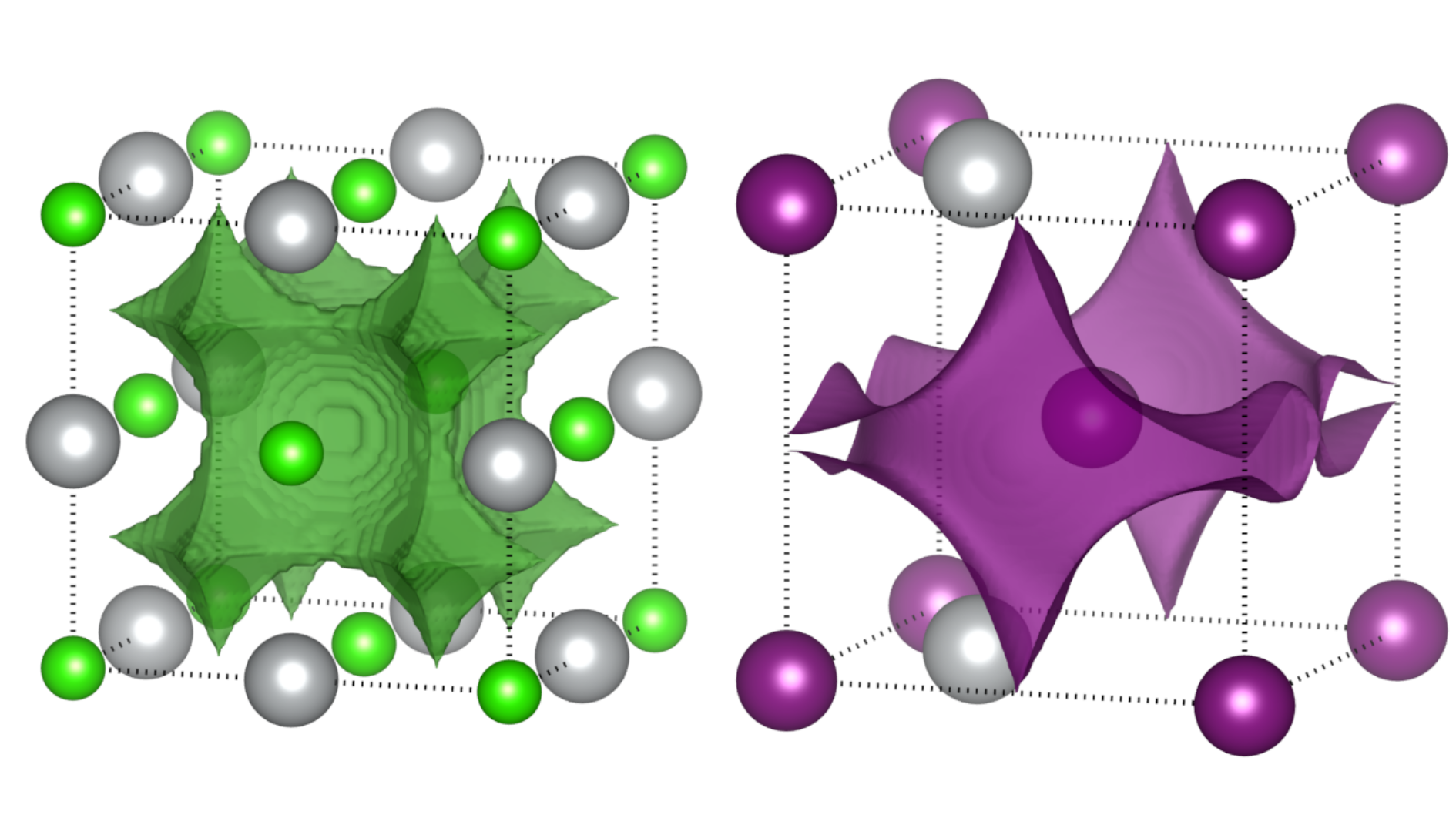}
    \caption{Three-dimensional potential energy profile, $U(\textbf{r})$, of a mobile Ag ion for AgCl (left panel) and $\alpha$-AgI (right panel). The isosurfaces show the minimum energy at which a continuous connecting pathway exists for the unit cell. The positions of the fixed ions are indicated. Ag atoms are represented in gray, while Cl and I atoms are represented in green and purple, respectively.}
    \label{fig:potential_map}
\end{figure}

The variations of the potential energy surface can be quantified by computing $U(\mathbf{r})$ from first principles, which we do, as illustration, for {\color{black} single unit cells of} the metal-halide electrolytes AgCl, LiCl, LiI, $\alpha$-AgI, and $\alpha$-CuBr, as shown in Fig.~\ref{fig:potential_map} for AgCl and $\alpha$-AgI. From here, we can obtain $\mu$ per mobile ion (Eq.~\ref{eqn:mu}), for each compound via the calculation of $\gamma(r)$, as defined in Eq.~\eqref{eqn:gamma_Hessian}. These ionic mobilities, $\mu_{calc}$, assuming $q=e$, are listed in Table~\ref{tab:mobilities}. A precise comparison between experiment and theory is not possible because of the lack of data for disparate samples taken under consistent experimental conditions. Moreover the presence of non-idealities in experimental samples - for instance, experimental samples are often polycrystalline, may exhibit size effects, and may have more than one mobile defect or ionic species - necessarily means that our calculated mobilities are not perfectly reflective of real-world measurements. Nonetheless, it is still useful to compare our results with ionic mobilities extracted from experimental studies of AgCl~\citep{Maier1988}, LiI~\citep{Poulsen1980}, $\alpha$-AgI~\citep{Sunandana2004}, and $\alpha$-CuBr~\citenum{Funke2013}. The ionic mobilities obtained by fitting the conductivity (see Methods section for details) are 0.08, 0.13, 0.022, and 0.001 $\mathrm{cm^2/V~s}$, respectively. {\color{black} While there are significant discrepancies between the calculated and theoretical results, the comparison of these values nevertheless shows a consistency within 1-2 orders of magnitude. After factoring in the experimental complexities, mentioned above, and the simplifying approximations used in the calculations, we believe that our results represent a promising first step toward reliable first principles determination of ionic mobility via Eq.~\eqref{eqn:mu}, and provide motivation for future experiments on clean, single-crystal samples under consistent conditions, whose measurements could be more readily compared with our simulated mobilities.} 

In comparison with more established methods such as deriving the conductivity from molecular dynamics simulations of diffusion, this technique has the advantage of being applicable to systems that have mobilities of any order of magnitude. In contrast, molecular dynamics are often limited to higher temperatures where enough diffusion events can be observed\cite{qi2021bridging,he2018statistical}.
On the other hand, in comparison with NEB methods, which allow us to explore only the minimum energy path, the method introduced here takes into account the whole potential landscape, and it is easily extendable to anisotropic systems.

\begin{table}[]
    \centering
    \begin{ruledtabular}
    \begin{tabular}{cccccc}
     compound & AgCl & LiCl & LiI & $\alpha$-AgI & $\alpha$-CuBr \\
     \hline
     $\mu_{calc}$ ($\mathrm{cm^{2}/V~s}$)  & 0.0012 & 0.043 & 0.99 & 0.25 & 0.18 \\
    \end{tabular}
    \end{ruledtabular}
    \caption{Directionally-averaged calculated ionic mobilities per mobile ion, $\mu_{calc}$ (Eq.~\ref{eqn:mu}), computed via $\gamma$ (Eq.~\eqref{eqn:gamma_Hessian}), using the potential from \textit{ab initio} calculations.}
    \label{tab:mobilities}
\end{table}

\section{Conclusions}
\label{sec:Conclusions}

In summary, we have developed a microscopic theory for ionic motion in crystals. We found that the ionic mobility depends essentially on the lattice softness (via the third power of the sound velocity) and the curvature of the atomic potential felt by the ions; namely, hard materials with smooth atomic potentials are the best candidates for high ionic mobility. This theory yields a tractable route for the calculation of ionic mobilities via modern \textit{ab initio} or other theoretical methods.  Further, the \textit{ab initio} approach can, in principle, be extended to account for the extrinsic effects that impact measured ionic mobilities, such as grain boundaries, impurities, and other types of defect that are already well-known in solid-state physics. Our numerical results represent a promising first step in the calculation of ionic mobilities from condensed matter perspective and without the use of molecular dynamics. The consistency of our results with experimentally extracted mobilities, while not wholly satisfactory, is encouraging and motivates both improved numerical calculations as well as new, well-controlled experiments that would allow for a true like-for-like comparison between theory and real-world measurements.

The last century has seen the development of a powerful theoretical framework to study the effect of defects and interfaces in the motion of electrons in solids. The same, however, cannot be said for the case of ions. The understanding of how ions interact with defects and interfaces in solids is an unexplored landscape, and any further progress in the development of solid-state electrolytes, which are the key elements of solid-state batteries, depends fundamentally on progress in this area of research.   

\acknowledgements

We acknowledge the National Research Foundation, Prime Minister Office, Singapore, under its Medium Sized Centre Programme. A.R. thanks the support by Yale-NUS College (through Grant No. A-0003356-42-00). The computational work was supported by the Centre of Advanced 2D Materials, funded by the National Research Foundation, Prime Minister's Office, Singapore, under its Medium-Sized Centre Programme.

\section*{Methods}
\paragraph{Density functional theory at 0~K --}
DFT calculations are performed using the Quantum \textsc{ESPRESSO}~\citep{Giannozzi2009,Giannozzi2017} code. Structural relaxations and total energy calculations are performed using a PAW basis~\citep{Blochl1994,DalCorso2014} and the Perdew-Burke-Ernzerhof (PBE) exchange-correlation functional~\citep{Perdew1996}. The kinetic energy cutoffs of the charge density and wavefunctions are set to at least the minimum recommended values of the PAW pseudopotential~\citep{DalCorso2014}. The Brillouin zones for all materials are sampled using uniform grids of $4\times4\times4$ (total energies) and $6\times6\times6$  (phonons) K-points.  

For the calculation of $U(r)$ we allow one ion of the mobile metal species to move while keeping all other ions fixed. The mobile ion is moved within the cubic unit cell by intervals of $1/64$ of the lattice parameter, $a$. Only configurations in which the distance from the mobile ion to any fixed ion is greater than ($5/12$)$a$ ($\alpha$-AgI and $\alpha$-CuBr) or ($1/3$)$a$ (AgCl, LiCl, LiI) are permitted. In the cases of $\alpha$-AgI and $\alpha$-CuBr we note that the metal ions have partial occupation and thus multiple possible positions. We therefore compute the total energies of all possible permutations of the positions of the mobile ions within the unit cell. The resulting minimum energy configurations are those in which the metal ions are located at the tetrahedral positions on adjacent faces, in agreement with AIMD calculations for $\alpha$-AgI~\citep{Wood2006}. Accordingly, to compute $U(r)$ for these materials we fix one of the tetrahedral metal ions and allow the other to move to all other permitted positions, as described above. 

Phonon calculations are performed using SG15 optimized norm-conserving Vanderbilt (ONCV) pseudopotentials~\citep{Hamann2013,Schlipf2015} and a PBE exchange-correlation functional, with a 60~Ry kinetic energy cutoff for wavefunctions. Sound velocities are derived from the phonon dispersions along the $\Gamma-X$ path for cubic cells, as defined in Ref.~\citenum{Setyawan2010}.

The volumetric images shown in Fig.~\ref{fig:potential_map} were generated in VESTA~\citep{Momma2011}.

\paragraph{\textit{Ab initio} molecular dynamics simulations --}
AIMD simulations are carried out using the SIESTA code~\citep{Soler2002}.
The forces are calculated using the local density approximation (LDA) of density functional theory~\citep{Ceperley1980}, and a Harris functional is used for the first step of the self-consistency cycle. 
The core electrons are represented by pseudopotentials of the Troullier-Martins scheme~\citep{Troullier1991}. 
The basis sets for the Kohn-Sham states are linear combinations of numerical atomic orbitals, of the polarized double-zeta type~\citep{Sanchez-Portal1997,Sanchez-Portal2001}.  
The $\Gamma$-point is used for Brillouin zone sampling. 
$\alpha$-AgI AIMD calculations are performed in 256 atom supercells. AIMD calculations for rocksalt structures are performed in 216 atom supercells. 
The temperature is controlled by means of a Nos\'{e} thermostat~\citep{Nose1984}. The integration time step used is 1~fs and the total integration time is 26~ps. The equilibration time varies between different temperatures and systems and is determined by examining the mean square displacement. 

\paragraph*{Calculations of the ionic mobility from existing experimental data --}
Experimental mobilities are found by fitting existing experimental data in the literature. We have assumed that there is only one mobile ion species. The conductivity is fitted using $\sigma=qn\mu$, where $n=N\exp\left(-E_{a}/k_BT\right)$ is the number of mobile ions. Here, $N$ is the total number of atoms of the mobile species in the crystal, $E_{a}$ is the activation energy necessary to make the ion mobile, $k_B$ is the Boltzmann constant and $T$ is the temperature.

\onecolumngrid
\appendix

\section{Matrix Elements}
\label{sec:Matrix_Elements}

Starting with $\langle \mathbf{r}_+^{j+1}, \mathbf{s}_+^{j+1}| e^{-\frac{i\hat{H}_i\Delta}{\hbar}}|\mathbf{r}_+^j,\mathbf{s}_+^j\rangle$, we note that the Hamiltonian is normal-ordered with respect to the second-quantization operators. This means that $a_s$ and $a_s^\dagger$ are replaced by $s_+^j$ and $\bar{s}_+^{j + 1}$ (since the annihilation operators act on the ket and creation ones act on the bra, they pick up the corresponding time slice index). This gives
\begin{align}
    &\langle \mathbf{r}_+^{j+1}, \mathbf{s}_+^{j+1}| e^{-\frac{i\hat{H}_i\Delta}{\hbar}}|\mathbf{r}_+^j,\mathbf{s}_+^j\rangle
    =\exp\left[-\frac{i\Delta}{\hbar}\sum_s \hbar \Omega_s\bar{s}_{+}^{j+1}s_{+}^j\right]
     \nonumber
     \\
     \times &
     \langle \mathbf{s}_+^{j+1}|\otimes\langle \mathbf{r}_+^{j+1}|
     \exp\left[-\frac{i\Delta}{\hbar}\left(
     \frac{1}{2}
    \hat{\mathbf{p}}^\dagger \mathbf{M}^{-1}\hat{\mathbf{p}} + U(\hat{\mathbf{r}},t)
     +\sum_s\left[\bar{s}_{+}^{j+1} C_s(\hat{\mathbf{r}},t) + s_+^jC_s(\hat{\mathbf{r}},t)\right]
     \right)\right]
     |\mathbf{r}_+^j\rangle\otimes|\mathbf{s}_+^j\rangle\,.
\end{align}
With all the second-quantization operators replaced by complex numbers, we can evaluate $\langle \mathbf{s}_+^{j+1}|\mathbf{s}_+^j\rangle = e^{\bar{\mathbf{s}}_+^{j+1}\mathbf{s}_+^{j}}$. We combine the exponential in the first line of the equation above with this term to get
\begin{equation}
    \exp\left[-\frac{i\Delta}{\hbar}\sum_s \hbar \Omega_s\bar{s}_{+}^{j+1}s_{+}^j\right] e^{\bar{\mathbf{s}}_+^{j+1}\mathbf{s}_+^{j}}
     = 
     \exp\left[-\frac{i\Delta}{\hbar}\sum_s \hbar \Omega_s\bar{s}_{+}^{j+1}s_{+}^j+\sum_s \bar{s}_+^{j+1}s_+^j\right]
     =
     \exp\left[\sum_s e^{-\frac{i\Delta\Omega_s}{\hbar}} \bar{s}_+^{j+1}s_+^j\right]\,,
\end{equation}
where the last equality holds because $\Delta\ll 1$.

Next, in the limit $\Delta\rightarrow 0$, the exponential can be split into four parts:
\begin{align}
    & \exp\left[-\frac{i\Delta}{\hbar}\left(
     \frac{1}{2}
    \hat{\mathbf{p}}^\dagger \mathbf{M}^{-1}\hat{\mathbf{p}} + U(\hat{\mathbf{r}},t)
     +\sum_s\left[\bar{s}_{+}^{j+1} C_s(\hat{\mathbf{r}},t) + s_+^jC_s(\hat{\mathbf{r}},t)\right]
     \right)\right]
     \nonumber
     \\
     \approx &
     \exp\left[-\frac{i\Delta}{\hbar}\sum_s\bar{s}_{+}^{j+1} C_s(\hat{\mathbf{r}},t) \right]
     \exp\left[-\frac{i\Delta}{\hbar}
     \frac{1}{2}
    \hat{\mathbf{p}}^\dagger \mathbf{M}^{-1}\hat{\mathbf{p}} \right]
     \exp\left[-\frac{i\Delta}{\hbar} U(\hat{\mathbf{r}},t)
     \right]
     \exp\left[-\frac{i\Delta}{\hbar}\sum_ss_+^j C_s(\hat{\mathbf{r}},t)\right]
     \,.
\end{align}
The first exponential acts on the bra so that $\hat{\mathbf{r}}\rightarrow\mathbf{r}^{j+1}_+$, while the last two act on the ket with $\hat{\mathbf{r}}\rightarrow\mathbf{r}^{j}_+$. The remaining part is
\begin{align}
     &
     \langle \mathbf{r}_+^{j+1}|
     e^{-\frac{i\Delta}{\hbar}\frac{1}{2}
    \hat{\mathbf{p}}^\dagger \mathbf{M}^{-1}\hat{\mathbf{p}}}
     |\mathbf{r}_+^j\rangle
     =
     \int d\mathbf{p} \langle \mathbf{r}_+^{j+1}|\mathbf{p}\rangle \langle \mathbf{p}|e^{-\frac{i\Delta}{\hbar}\frac{1}{2}
    \hat{\mathbf{p}}^\dagger \mathbf{M}^{-1}\hat{\mathbf{p}}}|\mathbf{r}_+^j\rangle
     =
     \int d\mathbf{p} e^{-\frac{i\Delta}{\hbar}\frac{1}{2}
    \mathbf{p}^T \mathbf{M}^{-1}\mathbf{p}}\langle \mathbf{r}_+^{j+1}|\mathbf{p}\rangle \langle \mathbf{p}|\mathbf{r}_+^j\rangle
     \\
     =&
     \int d\mathbf{p}
    \exp\left[-\frac{i\Delta}{\hbar}\frac{1}{2}
    \mathbf{p}^T \mathbf{M}^{-1}\mathbf{p}+i\mathbf{p}^T\frac{\mathbf{r}_+^{j+1} - \mathbf{r}_+^j}{\hbar}\right]
    \frac{1}{\left(2\pi \hbar\right)^{ID}}
    =
    \left|\frac{\mathbf{M}}{2\pi i \Delta \hbar}\right|^{1/2}
    \exp\left[\frac{i\left(\mathbf{r}_+^{j+1} - \mathbf{r}_+^j\right)^T\mathbf{M}\left(\mathbf{r}_+^{j+1} - \mathbf{r}_+^j\right)}{2\Delta\hbar}\right] \,,
    \nonumber
\end{align}
where $I$ is the number of mobile ions. Combining the components gives
\begin{align}
        &\langle \mathbf{r}_+^{j+1}, \mathbf{s}_+^{j+1}| e^{-\frac{i\hat{H}_i\Delta}{\hbar}}|\mathbf{r}_+^j,\mathbf{s}_+^j\rangle   = 
     \exp\left[\sum_s e^{-\frac{i\Delta\Omega_s}{\hbar}} \bar{s}_+^{j+1}s_+^j\right]
     \left|\frac{\mathbf{M}}{2\pi i \Delta \hbar}\right|^{1/2}
    \nonumber
    \\
     \times &
    \exp\left[-\frac{i\Delta}{\hbar}
     \sum_s\left[\bar{s}_{+}^{j+1}C_s(\mathbf{r}_+^{j+1},(j+1)\Delta) + s_{+}^jC_s(\mathbf{r}_+^j,j\Delta)\right]
    \right]
    \nonumber
    \\
    \times &\exp\left[\frac{i\left(\mathbf{r}_+^{j+1} - \mathbf{r}_+^j\right)^T\mathbf{M}\left(\mathbf{r}_+^{j+1} - \mathbf{r}_+^j\right)}{2\Delta\hbar}-\frac{i\Delta}{\hbar}U(\mathbf{r}_+^j,j\Delta)\right] \,,
    \label{eqn:matrix_el_1}
\end{align}

Similar steps lead to 
\begin{align}
    & \langle \mathbf{r}_-^{j}, \mathbf{s}_-^{j}| e^{\frac{i\hat{H}_i\Delta}{\hbar}}|\mathbf{r}_-^{j+1},\mathbf{s}_-^{j+1}\rangle=
    \exp\left[\sum_s e^{\frac{i\Delta\Omega_s}{\hbar}} \bar{s}_-^{j}s_-^{j+1}\right]\left|\frac{i\mathbf{M}}{2\pi \Delta \hbar}\right|^{1/2}
    \nonumber
    \\
    \times &
    \exp\left[\frac{i\Delta}{\hbar}
     \sum_s\left[\bar{s}_{-}^{j}C_s(\mathbf{r}_-^{j}, j\Delta) + s_{-}^{j+1}C_s(\mathbf{r}_-^{j+1}, (j + 1)\Delta)\right]
    \right]
    \nonumber
    \\
    \times&
    \exp\left[-\frac{i\left(\mathbf{r}_-^{j+1} - \mathbf{r}_-^j\right)^T\mathbf{M}\left(\mathbf{r}_-^{j+1} - \mathbf{r}_-^j\right)}{2\Delta\hbar}+\frac{i\Delta}{\hbar}U(\mathbf{r}_-^j, j\Delta)\right]\,.
    \label{eqn:matrix_el_2}
\end{align}
A quick way to do it is to replace $\Delta\rightarrow -\Delta$, switch the subscripts from $+$ to $-$, and interchange $j \leftrightarrow j+ 1$.

For $\langle \mathbf{r}^{N}_-, \mathbf{s}^{N}_-|\hat{O}| \mathbf{r}^N_+,\mathbf{s}^N_+ \rangle$, we have
\begin{align}
    &\left(\langle \mathbf{r}^{N}_-|\otimes\langle \mathbf{s}^{N}_-|\right)\left(\hat{O}\otimes \hat{1}\right) \left(| \mathbf{r}^N_+\rangle\otimes |\mathbf{s}^N_+ \rangle\right)
    =
    \langle \mathbf{r}^{N}_-|\hat{O}|\mathbf{r}^N_+\rangle \langle\mathbf{s}^{N}_-|\mathbf{s}^{N}_+\rangle = \langle \mathbf{r}^{N}_-|\hat{O}|\mathbf{r}^N_+\rangle e^{\bar{\mathbf{s}}^{N}_-\mathbf{s}^N_+}\,.
     \label{eqn:matrix_el_3}
\end{align}

Finally, we calculate $\langle \mathbf{r}^1_+, \mathbf{s}^1_+|\hat{\rho}_0| \mathbf{r}^{1}_-,\mathbf{s}^{1}_- \rangle$. We assume that at $t = 0$ the mobile and stationary ions and described by their own independent density operators. Specifically, we allow the stationary ions to be in a thermal equilibrium with an external bath, while the mobile ions start with a known density distribution $\hat{\rho}_m$, allowing us to write $\hat{\rho}_0 = \hat{\rho}_m\otimes e^{-\beta\hat{H}_S}$, so that
\begin{align}
    &\langle \mathbf{r}^1_+, \mathbf{s}^1_+|\hat{\rho}_0| \mathbf{r}^{1}_- ,\mathbf{s}^{1}_- \rangle
    =
    \langle \mathbf{r}^1_+|\hat{\rho}_m|\mathbf{r}^{1}_-\rangle  \langle \mathbf{s}^1_+| e^{-\beta\hat{H}_S} |\mathbf{s}^{1}_- \rangle
    =
    \langle \mathbf{r}^1_+|\hat{\rho}_m|\mathbf{r}^{1}_-\rangle  
    \exp\left(\sum_se^{-\beta\hbar\Omega_s}\bar{s}_{+}s_{-}^1\right)\,,
    \label{eqn:matrix_el_4}
\end{align}
where we have used $\langle \phi|e^{k a^\dagger a}|\psi\rangle = e^{e^k\bar{\phi}\psi}$.

\section{Field Integration}
\label{sec:Field_Integration}

Plugging Eqs.~\eqref{eqn:matrix_el_1}-\eqref{eqn:matrix_el_4} into Eq.~\eqref{eqn:O_tau} gives
\begin{align}
    \langle \hat{O}\rangle (\tau) 
    &=
    \frac{1}{\mathrm{Tr}\left[\hat{\rho}_0\right]}
    \int \mathcal{D}\left(\dots\right)
    \langle \mathbf{r}_-^N|\hat{O}|\mathbf{r}_+^N\rangle \langle \mathbf{r}^1_+|\hat{\rho}_m\mathbf{r}^{1}_-\rangle
    \left|\frac{\mathbf{M}}{2\pi  \Delta \hbar}\right|^{N-1}
    \nonumber
    \\
    &\times
    \prod_s
    \exp\left[-\sum_{j = 1}^N \left(
    \bar{s}_-^j s_-^j 
    + 
    \bar{s}_+^j s_+^j 
    \right)+
    \bar{s}_+^1 s_-^1e^{-\beta\hbar\Omega_s} + \bar{s}_-^N s_+^N\right]
    \nonumber
    \\
    &\times
    \prod_s 
    \exp\left[
    \sum_{j = 1}^{N - 1}e^{-i\Delta \Omega_s} \bar{s}_{+}^{j+1} s_{+}^j
    -
    \frac{i\Delta}{\hbar}
     \left[\bar{s}_{+}^{j+1}C_s(\mathbf{r}_+^{j+1}) + s_{+}^jC_s(\mathbf{r}_+^j)\right]
    \right]
    \nonumber
    \\
    &\times 
    \prod_s
    \exp\left[
    \sum_{j = 1}^{N - 1}
    e^{i\Delta  \Omega_s} \bar{s}_-^{j}s_-^{j+1} +\frac{i\Delta}{\hbar}
     \left[\bar{s}_-^{j}C_s(\mathbf{r}_-^{j}) + s_-^{j+1}C_s(\mathbf{r}_-^{j+1})\right]
    \right]
    \nonumber
    \\
    & \times 
     \prod_{j = 1}^{N - 1}
    \exp\left[\sum_{\sigma = \pm}\sigma\frac{i\left(\mathbf{r}_\sigma^{j+1} - \mathbf{r}_\sigma^j\right)^T\mathbf{M}\left(\mathbf{r}_\sigma^{j+1} - \mathbf{r}_\sigma^j\right)}{2\Delta\hbar}-\sigma\frac{i\Delta}{\hbar} U(\mathbf{r}_\sigma^j)
    \right]\,,
    \label{eqn:O_tau_matrix}
\end{align}
where $\hat{\rho}_m$ is the density operator for the mobile ions at $t = 0$. Note that we suppress the redundant time label in $C_s$ and $U$ because it is already present as the superscript of $\mathbf{r}$.

The next step involves integrating over the complex numbers $s_\pm^j$ and $\bar{s}_\pm^j$. Before we do that, however, there are two important features worth highlighting in Eq.~\eqref{eqn:O_tau_matrix}. First, each vibrational mode $s$ has $4N$ complex variables associated with it: $s_\pm^j$ and $\bar{s}_\pm^j$ for $j = 1\dots N$. Second, the different modes do not mix directly as there are no products of the form $\bar{s}_\pm^j s'^k_\pm$. This feature considerably simplifies the integration.

Picking out only the terms in the second, third, and fourth lines of Eq.~\eqref{eqn:O_tau_matrix} that depend on the mode $s$ allows us to define a multidimensional complex Gaussian integral
\begin{align}
    \mathcal{I}_s &= \int\mathcal{D}\left(\dots\right)
    \exp\left[
    \bar{s}_+^1 s_-^1e^{-\beta\hbar\Omega_s} + \bar{s}_-^N s_+^N
    \right]
    \nonumber
    \\
    &\times
     \exp\Bigg[-\sum_{j = 1}^N \left(
    \bar{s}_-^j s_-^j 
    + 
    \bar{s}_+^j s_+^j 
    \right)
    \Bigg]
    \nonumber
        \\
    &\times
     \exp\Bigg[
    \sum_{j = 1}^{N - 1}e^{-i\Delta \Omega_s}\bar{s}_{+}^{j+1} s_{+}^j
    +
    \sum_{j = 1}^{N - 1}
    e^{i\Delta  \Omega_s}\bar{s}_-^{j}s_-^{j+1}\Bigg]
    \nonumber
    \\
    &\times
    \exp\Bigg[
    -\sum_{j = 1}^{N - 1}
    \frac{i\Delta}{\hbar}
     \left[\bar{s}_{+}^{j+1}C_s(\mathbf{r}_+^{j+1}) + s_{+}^jC_s(\mathbf{r}_+^j)\right]
    \Bigg]
    \nonumber
    \\
    &\times
    \exp\Bigg[
    \sum_{j = 1}^{N - 1}
   \frac{i\Delta}{\hbar}
     \left[\bar{s}_-^{j}C_s(\mathbf{r}_-^{j}) + s_-^{j+1}C_s(\mathbf{r}_-^{j+1})\right]
    \Bigg]
    \,.
    \label{eqn:I}
\end{align}

To evaluate the integral in Eq.~\eqref{eqn:I}, we first write the terms inside the exponential as
\begin{align}
    &
    -\begin{pmatrix}
        \bar{s}_+ & \bar{s}_-
    \end{pmatrix}
    \left(
    \begin{array}{ccc|ccc}
        1 & 0 & \dots & \dots & 0 & -e^{-\beta\hbar\Omega_s}
        \\
        -e^{-i\Delta\Omega_s} & 1 & \dots & \dots & 0 & 0
        \\
        \vdots& \vdots & \ddots & \iddots & \vdots & \vdots
        \\
        \hline
        \dots & 0 & -1 &  1 & 0 & \dots
        \\
        \dots & 0 & 0 &  -e^{i\Delta\Omega_s} & 1 & \dots
        \\
        \iddots & \vdots & \vdots &  \vdots & \vdots & \ddots
    \end{array}
    \right)
    \begin{pmatrix}
         s_+ \\  s_-
    \end{pmatrix}
    -
    \frac{i\Delta}{\hbar}\begin{pmatrix}
        \bar{s}_+ & \bar{s}_-
    \end{pmatrix}
    \begin{pmatrix}
    0\\C_s(\mathbf{r}_+^2)\\\vdots\\C_s(\mathbf{r}_+^N)\\0\\-C_s(\mathbf{r}^{N-1}_-) \\ \vdots \\ -C_s(\mathbf{r}^{1}_-)
    \end{pmatrix}
    \label{eqn:Gaussian_matrix}
    \\
    &-
    \frac{i\Delta}{\hbar}
    \begin{pmatrix}
    C_s(\mathbf{r}_+^1) & \dots & C_s(\mathbf{r}_+^{N-1}) & 0 & -C_s(\mathbf{r}_-^N)&\dots&-C_s(\mathbf{r}_-^2)&0
    \end{pmatrix}
    \begin{pmatrix}
         s_+ \\  s_-
    \end{pmatrix}\,,
    \nonumber
\end{align}
where $\begin{pmatrix}\bar{s}_+ &\bar{s}_- \end{pmatrix} = \begin{pmatrix}\bar{s}_+^1 & \bar{s}_+^2&\dots&\bar{s}_+^N&\bar{s}_-^N & \dots & \bar{s}^2_- & \bar{s}^1_- \end{pmatrix}$.

Performing the integral requires inverting the matrix in Eq.~\eqref{eqn:Gaussian_matrix}. The top left (bottom right) quadrants of this matrix are lower bidiagonal matrices with 1 on the main diagonal and $-e^{-i\Delta\Omega_s}$ ($-e^{i\Delta\Omega_s}$) on the first subdiagonal. The remaining two quadrants have a single non-zero entry each, located at their top right corner. It is convenient to write the inverse as
\begin{align}
    G_s &= 
    \begin{pmatrix}
        G^{++}_s & G^{+-}_s
        \\
        G^{-+}_s & G^{--}_s
    \end{pmatrix}
    \nonumber
    \\
    \left[G_s^{++}\right]_{ln} &=
    e^{-i\Delta \Omega_s ( l - n)}\left[\Theta(l-n) +  n_B\left(\Omega_s\right)\right]
    \nonumber
    \\
    \left[G_s^{--}\right]_{ln} &=
    e^{i\Delta \Omega_s ( l - n)}\left[\Theta(l-n) +   n_B\left(\Omega_s\right)\right]
    \nonumber
    \\
    \left[G_s^{+-}\right]_{ln} &= e^{i\Delta\Omega_s(N+1 - l - n)}n_B\left(\Omega_s\right)
    \nonumber
    \\
    \left[G_s^{-+}\right]_{ln} &= e^{-i\Delta\Omega_s(N+1 - l - n)}\left[n_B\left(\Omega_s\right)+1\right]\,,
    \label{eqn:Inverted_Matrix}
\end{align}
where the discrete Heaviside function $\Theta(0) = 1$ and $n_B(x)$ is the Bose-Einstein distribution. The details of the inversion procedure can be found in Appendix~\ref{sec:Matrix_Inversion}. The resulting expression is
\begin{align}
    \mathcal{I}_s &= \left|G_s\right|
    \exp\left[-\frac{\Delta^2}{\hbar^2}\sum_{ln = 1}^N 
    C_s(\mathbf{r}^l_+)
    \left[G_s^{++}\right]_{ln}
    C_s(\mathbf{r}_+^n)
    \left(1-\delta_{n,1}\right)\left(1-\delta_{l,N}\right)
   \right]
    \nonumber
    \\
    &\times
    \exp\left[
    -\frac{\Delta^2}{\hbar^2}\sum_{ln = 1}^N 
    C_s(\mathbf{r}^{N + 1-l}_-)
    \left[G_s^{--}\right]_{ln}
    C_s(\mathbf{r}_-^{N+1 - n})
    \left(1-\delta_{n,1}\right)\left(1-\delta_{l,N}\right)
    \right]
    \nonumber
    \\
    &\times\exp\left[\frac{\Delta^2}{\hbar^2}\sum_{ln = 1}^N 
    C_s(\mathbf{r}^{l}_+)
    \left[G_s^{+-}\right]_{ln}
    C_s(\mathbf{r}_-^{N+1 - n})
    \left(1-\delta_{n,1}\right)\left(1-\delta_{l,N}\right)
    \right]
    \nonumber
    \\
    &\times\exp\left[\frac{\Delta^2}{\hbar^2}\sum_{ln = 1}^N 
    C_s(\mathbf{r}^{N+1-l}_-)
    \left[G_s^{-+}\right]_{ln}
    C_s(\mathbf{r}_+^n)
    \left(1-\delta_{n,1}\right)\left(1-\delta_{l,N}\right)
    \right]\,.
    \label{eqn:I_2}
\end{align}
Here, the $\pi^{2N}$ term from the Gaussian integration is cancelled by $\pi^{2N}$ in the denominator originating from Eq.~\eqref{eqn:Identity}. The expression can be made more symmetric by relabeling $N + 1 - n\rightarrow n$ for the $\mathbf{r}_-^{N + 1 -n}$ terms (and, of course, changing the corresponding index of the $G$ matrix element). In addition, for the sake of brevity, we will suppress the Kronecker deltas and set $C_s\left(\mathbf{r}_+^{l = N}\right)=C_s\left(\mathbf{r}_-^{l = 1}\right) = C_s\left(\mathbf{r}_+^{n = 1}\right)=C_s\left(\mathbf{r}_-^{n = N}\right) = 0$ implicitly. Plugging in the expressions for the inverse matrix elements yields
\begin{align}
    \mathcal{I}_s &= \left|G_s\right|
    \nonumber
    \\
    &\times
    \exp\left[-\frac{\Delta^2}{\hbar^2}\sum_{ln = 1}^N 
    C_s(\mathbf{r}^l_+)
    e^{-i\Delta \Omega_s ( l - n)}\left[\Theta(l-n) + n_B\left(\Omega_s\right)\right]
    C_s(\mathbf{r}_+^n)
   \right]
    \nonumber
    \\
    &\times
    \exp\left[
    -\frac{\Delta^2}{\hbar^2}\sum_{ln = 1}^N 
    C_s(\mathbf{r}^{l}_-)
    e^{i\Delta \Omega_s ( -l + n)}\left[\Theta(n-l) + n_B\left(\Omega_s\right)\right]
    C_s(\mathbf{r}_-^{n})
    \right]
    \nonumber
    \\
    &\times\exp\left[\frac{\Delta^2}{\hbar^2}\sum_{ln = 1}^N 
    C_s(\mathbf{r}^{l}_+)
    e^{i\Delta\Omega_s(n - l)} n_B\left(\Omega_s\right)
    C_s(\mathbf{r}_-^{n})
    \right]
    \nonumber
    \\
    &\times\exp\left[\frac{\Delta^2}{\hbar^2}\sum_{ln = 1}^N 
    C_s(\mathbf{r}^{l}_-)
    e^{-i\Delta\Omega_s(l - n)}\left[n_B\left(\Omega_s\right)+1\right]
    C_s(\mathbf{r}_+^n)
    \right]\,.
    \label{eqn:I_3}
\end{align}
Note that the phase term $e^{-i\Delta\Omega_s(l - n)}$ in Eq.~\eqref{eqn:I_3} is the same for all the exponentials, allowing us to combine the terms as
\begin{align}
    &
    \begin{pmatrix}
    C_s(\mathbf{r}_+^l) & C_s(\mathbf{r}_-^l)
    \end{pmatrix}
    \begin{pmatrix}
    \Theta(l-n) + n_B\left(\Omega_s\right) & -n_B\left(\Omega_s\right)
    \\
    -1 -n_B\left(\Omega_s\right)& \Theta(n-l) + n_B\left(\Omega_s\right)
    \end{pmatrix}
    \begin{pmatrix}
    C_s(\mathbf{r}_+^n)) \\ C_s(\mathbf{r}_-^n)
    \end{pmatrix}
    \nonumber
    \\
    =&
    \frac{1}{\sqrt{2}}
    \begin{pmatrix}
    C_s(\mathbf{r}_+^l) + C_s(\mathbf{r}_-^l) & C_s(\mathbf{r}_+^l) - C_s(\mathbf{r}_-^l)
    \end{pmatrix}
    \begin{pmatrix}
    \delta_{ln}/2 & -\Theta(n - l)
    \\
    \Theta(l - n)& 1+2 n_B\left(\Omega_s\right) + \delta_{ln}/2
    \end{pmatrix}
    \begin{pmatrix}
    C_s(\mathbf{r}_+^n) + C_s(\mathbf{r}_-^n)
    \\
    C_s(\mathbf{r}_+^n) - C_s(\mathbf{r}_-^n)
    \end{pmatrix}
    \frac{1}{\sqrt{2}}
    \nonumber
    \\
    \rightarrow &
    \frac{1}{\sqrt{2}}
    \begin{pmatrix}
    C_s(\mathbf{r}_+^l) + C_s(\mathbf{r}_-^l) & C_s(\mathbf{r}_+^l) - C_s(\mathbf{r}_-^l)
    \end{pmatrix}
    \begin{pmatrix}
    0 & -\Theta(n - l)
    \\
    \Theta(l - n)&\coth\left(\frac{\beta\hbar\Omega_s}{2}\right)
    \end{pmatrix}
    \begin{pmatrix}
    C_s(\mathbf{r}_+^n) + C_s(\mathbf{r}_-^n)
    \\
    C_s(\mathbf{r}_+^n) - C_s(\mathbf{r}_-^n)
    \end{pmatrix}
    \frac{1}{\sqrt{2}}\,.
\end{align}
We drop the $\delta_{ln}$ terms because their contribution decays as $\sim 1 / N$: $\Delta^2\propto N^{-2}$ in the prefactor, while the Kronecker deltas provide only $N$ terms. 

Next, we write $\mathcal{I}_s$ as
\begin{align}
    \mathcal{I}_s &=\left|G_s\right|
    \exp\Bigg[-\frac{\Delta^2}{\hbar^2}\sum_{ln = 1}^N 
    \frac{e^{-i\Delta \Omega_s ( l - n)}}{2}
    \nonumber
    \\
    &\times\begin{pmatrix}
    C_s(\mathbf{r}_+^l) + C_s(\mathbf{r}_-^l) & C_s(\mathbf{r}_+^l) - C_s(\mathbf{r}_-^l)
    \end{pmatrix}
    \begin{pmatrix}
    0 & -\Theta(n - l)
    \\
    \Theta(l - n)& \coth\left(\frac{\beta\hbar\Omega_s}{2}\right)
    \end{pmatrix}
    \begin{pmatrix}
    C_s(\mathbf{r}_+^n) + C_s(\mathbf{r}_-^n)
    \\
    C_s(\mathbf{r}_+^n) - C_s(\mathbf{r}_-^n)
    \end{pmatrix}
    \Bigg]
    \nonumber
    \\
    &=\left|G_s\right|
    \exp\Bigg[\frac{\Delta^2}{\hbar^2}\sum_{ln = 1}^N 
    \frac{e^{-i\Delta \Omega_s ( l - n)}}{2}
    \left( C_s(\mathbf{r}_+^l) + C_s(\mathbf{r}_-^l)\right)
    \Theta(n - l)
    \left(C_s(\mathbf{r}_+^n) - C_s(\mathbf{r}_-^n)\right)
    \Bigg]
    \nonumber
    \\
    &\times \exp\Bigg[-\frac{\Delta^2}{\hbar^2}\sum_{ln = 1}^N 
    \frac{e^{-i\Delta \Omega_s ( l - n)}}{2}
    \left(C_s(\mathbf{r}_+^l) - C_s(\mathbf{r}_-^l)\right)
    \Theta(l - n)
    \left(
     C_s(\mathbf{r}_+^n) + C_s(\mathbf{r}_-^n)
     \right)
    \Bigg]
    \nonumber
    \\
    &\times \exp\Bigg[-\frac{\Delta^2}{\hbar^2}\sum_{ln = 1}^N 
    \frac{e^{-i\Delta \Omega_s ( l - n)}}{2}
    \left(C_s(\mathbf{r}_+^l) - C_s(\mathbf{r}_-^l)\right)
    \coth\left(\frac{\beta\hbar\Omega_s}{2}\right)
    \left(C_s(\mathbf{r}_+^n) - C_s(\mathbf{r}_-^n)\right)
    \Bigg]
    \nonumber
    \\
    &=\left|G_s\right|
    \exp\Bigg[i\frac{\Delta^2}{\hbar^2}\sum_{ln = 1}^N 
    \sin\left(\Delta\Omega_s(n - l)\right)
    \left(C_s(\mathbf{r}_+^n) - C_s(\mathbf{r}_-^n)\right)
    \Theta(n - l)
    \left(C_s(\mathbf{r}_+^l) + C_s(\mathbf{r}_-^l)\right)
    \Bigg]
    \nonumber
    \\
    &\times \exp\Bigg[-\frac{\Delta^2}{\hbar^2}\sum_{ln = 1}^N 
    \frac{e^{-i\Delta \Omega_s ( l - n)}}{2}
    \left(C_s(\mathbf{r}_+^l) - C_s(\mathbf{r}_-^l)\right)
    \coth\left(\frac{\beta\hbar\Omega_s}{2}\right)
    \left(C_s(\mathbf{r}_+^n) - C_s(\mathbf{r}_-^n)\right)
    \Bigg]\,,
\end{align}
where we drop the $C_s\left(\mathbf{r}_+^{l = N}\right)=C_s\left(\mathbf{r}_-^{l = 1}\right) = C_s\left(\mathbf{r}_+^{n = 1}\right)=C_s\left(\mathbf{r}_-^{n = N}\right) = 0$ requirement for $\Delta\rightarrow 0$ as the contribution of these terms goes as $\sim 1 / N$.

The term $|G_s| = |G_s^{-1}|^{-1} = 1/(1 - e^{-\beta\hbar\Omega_s}) = \mathrm{Tr} \left[e^{-\beta\hbar\Omega b_s^\dagger b_s}\right] = \mathrm{Tr} \left[\hat{\rho}_s\right]$ cancels $\mathrm{Tr}\left[\hat{\rho}_s\right]$ in the denominator of Eq.~\eqref{eqn:O_tau_matrix} for each mode $s$, leaving only $\mathrm{Tr}\left[\hat{\rho}_m\right]$. Combining all the $\mathcal{I}_s$ terms gives
\begin{align}
    \langle \hat{O}\rangle (\tau) 
    =&
    \frac{1}{\mathrm{Tr}\left[\hat{\rho}_m\right]}
    \int \mathcal{D}\left(\dots\right)
    \langle \mathbf{r}_-^N|\hat{O}|\mathbf{r}_+^N\rangle \langle \mathbf{r}^1_+|\hat{\rho}_m|\mathbf{r}^{1}_- \rangle
    \left|\frac{\mathbf{M}}{2\pi  \Delta \hbar}\right|^{N-1}
    \prod_s \exp\left[-\frac{1}{2}
    \coth\left(\frac{\beta\hbar\Omega_s}{2}\right)
    Q_sQ^*_s
    \right]
    \nonumber
    \\
    \times&
    \prod_s \exp\Bigg[i\frac{\Delta^2}{\hbar^2}\sum_{ln = 1}^N 
    \sin\left(\Delta\Omega_s(n - l)\right)
    \left(C_s(\mathbf{r}_+^n) - C_s(\mathbf{r}_-^n)\right)
    \Theta(n - l)
    \left( C_s(\mathbf{r}_+^l) + C_s(\mathbf{r}_-^l)\right)
    \Bigg]
    \nonumber
    \\
    \times &
    \prod_{j = 1}^{N - 1}
    \exp\left[\sum_{\sigma = \pm}\sigma\frac{i\left(\mathbf{r}_\sigma^{j+1} - \mathbf{r}_\sigma^j\right)^T\mathbf{M}\left(\mathbf{r}_\sigma^{j+1} - \mathbf{r}_\sigma^j\right)}{2\Delta\hbar}-\sigma\frac{i\Delta}{\hbar} U(\mathbf{r}_\sigma^j)
    \right]
    \nonumber
    \\
     =&
    \frac{1}{\mathrm{Tr}\left[\hat{\rho}_m\right]}
    \int \mathcal{D}\left(\dots\right)
    \langle \mathbf{r}_-^N|\hat{O}|\mathbf{r}_+^N\rangle \langle \mathbf{r}^1_+|\hat{\rho}_m|\mathbf{r}^{1}_- \rangle
    \left|\frac{\mathbf{M}}{2\pi  \Delta \hbar}\right|^{N-1}
    \prod_s \exp\left[-\frac{1}{2}
    \coth\left(\frac{\beta\hbar\Omega_s}{2}\right)
    Q_sQ^*_s
    \right]
    \nonumber
    \\
    \times&
    \prod_s \exp\Bigg[i\frac{\Delta^2}{\hbar^2}\sum_{ln = 1}^N 
    \sin\left(\Delta\Omega_s(n - l)\right)
    \left(Y_s(\mathbf{r}_+^n) - Y_s(\mathbf{r}_-^n)\right)
    \Theta(n - l)
    \left( Y_s(\mathbf{r}_+^l) + Y_s(\mathbf{r}_-^l) + 2W_s(l\Delta)\right)
    \Bigg]
    \nonumber
    \\
    \times &
    \prod_{j = 1}^{N - 1}
    \exp\left[\sum_{\sigma = \pm}\sigma\frac{i\left(\mathbf{r}_\sigma^{j+1} - \mathbf{r}_\sigma^j\right)^T\mathbf{M}\left(\mathbf{r}_\sigma^{j+1} - \mathbf{r}_\sigma^j\right)}{2\Delta\hbar}-\sigma\frac{i\Delta}{\hbar} U(\mathbf{r}_\sigma^j)
    \right]
    \label{eqn:O_tau_5}
\end{align}
with
\begin{equation}
     Q_s = \frac{\Delta}{\hbar}\sum_{l = 1}^N 
    e^{-i\Delta \Omega_s l}
    \left[
    C_s(\mathbf{r}_+^l) - C_s(\mathbf{r}_-^l)
    \right]
    = \frac{\Delta}{\hbar}\sum_{l = 1}^N 
    e^{-i\Delta \Omega_s l}
    \left[
    Y_s(\mathbf{r}_+^l) - Y_s(\mathbf{r}_-^l)
    \right]
    \,.
\end{equation}

\section{Inverting the Mode Matrix}
\label{sec:Matrix_Inversion}
Our goal is to invert
\begin{align}
    Y &= \left(
    \begin{array}{ccc|ccc}
        1 & 0 & \dots & \dots & 0 & -e^{-\beta\hbar\Omega_s}
        \\
        -e^{-i\theta} & 1 & \dots & \dots & 0 & 0
        \\
        \vdots& \vdots & \ddots & \iddots & \vdots & \vdots
        \\
        \hline
        \dots & 0 & -1 &  1 & 0 & \dots
        \\
        \dots & 0 & 0 &  -e^{i\theta} & 1 & \dots
        \\
        \iddots & \vdots & \vdots &  \vdots & \vdots & \ddots
    \end{array}
    \right)
    =
    \begin{pmatrix}
    a & b 
    \\
    c & d
    \end{pmatrix}\,.
\end{align}
Using the Banachiewicz identity,
\begin{equation}
    Y^{-1} = \begin{pmatrix}
   a^{-1} +a^{-1}b \left(d - c a^{-1}b\right)^{-1}ca^{-1}&-a^{-1}b\left(d - c a^{-1}b\right)^{-1}
    \\
    -\left(d - c a^{-1}b\right)^{-1}ca^{-1} & \left(d - c a^{-1}b\right)^{-1}
    \end{pmatrix}\,.
\end{equation}

The advantage here is that $a$ is a bidiagonal matrix with 1's on the main diagonal and identical entries on the subdiagonal. Writing $a = \mathbf{1} - S$, we have
\begin{equation}
    a^{-1} = (\mathbf{1} - S)^{-1} = \sum_{n = 0}^\infty S^n
\end{equation}
One can check that for $n > N$, $S^n$ vanishes while for $n \leq N$, the negative of the subdiagonal entry of $a$ is raised to the power $n$ and positioned on the $n$th diagonal. In other words, $[a^{-1}]_{jk} = \Theta(j - k)e^{-i(j - k)\theta}$, where we define Heaviside function $\Theta(0) =  1$.

Next, we have
\begin{align}
    \left[ca^{-1}b\right]_{jk} &= 
    \sum_{lm} c_{jl}a^{-1}_{lm}b_{mk} =
    \sum_{lm} (-1 \delta_{j,1}\delta_{l,N})a^{-1}_{lm}(-e^{-\beta\hbar\Omega_s} \delta_{m,1}\delta_{k,N})
    =
   a^{-1}_{N1}e^{-\beta\hbar\Omega_s}\delta_{j,1}\delta_{k,N}
   \nonumber
   \\
   &=
   e^{-i(N-1)\theta}e^{-\beta\hbar\Omega_s}\delta_{j,1}\delta_{k,N}\,,
   \\
   \left[a^{-1}b\right]_{jk} &= 
    \sum_{m} a^{-1}_{jm}b_{mk} =
    \sum_{m} a^{-1}_{jm}(-e^{-\beta\hbar\Omega_s} \delta_{m,1}\delta_{k,N})
    =
   -e^{-i(j - 1)\theta}e^{-\beta\hbar\Omega_s}\delta_{k,N}\,,
   \\
   \left[ca^{-1}\right]_{jk} &= 
    \sum_{l} c_{jl}a^{-1}_{lk} =
    \sum_{l} (-1 \delta_{j,1}\delta_{l,N})a^{-1}_{lk}
    =
    -\delta_{j,1}e^{-i(N - k)\theta}\,.
\end{align}

To obtain $\left(d - c a^{-1}b\right)^{-1}$, note that $d$ is also bidiagonal with $1$'s on the diagonal and identical terms on the sub-diagonal. Subtracting $c a^{-1}b$ adds a single element $-e^{-i(N-1)\theta}e^{-\beta\hbar\Omega_s}$ to the top right corner. Hence, we need to invert
\begin{equation}
    X = \left(d - c a^{-1}b\right) = \begin{pmatrix}
    1 & 0 & 0 & \dots &-e^{-i(N - 1)\theta}e^{-\beta\hbar\Omega_s}
    \\
    -e^{i\theta} & 1 & 0 &\dots&0 
    \\
    0&-e^{i\theta} & 1  &\dots&0 
    \\
    \vdots & \vdots &\vdots &\ddots & \vdots
    \end{pmatrix} =\begin{pmatrix}
    A & B 
    \\
    C & D
    \end{pmatrix} \,.
\end{equation}
Here, $D = 1$, $B$ is the last column of $X$ without the final elements, $C$ is the last row of $X$ without the last element, and $A$ is the remaining $(N - 1) \times (N-1)$ matrix. Invoking the Banachiewicz identity again, we write
\begin{align}
    \left(d - c a^{-1}b\right)^{-1} &= 
    \begin{pmatrix}
   A^{-1} +A^{-1}B \left(D - C A^{-1}B\right)^{-1}CA^{-1}&-A^{-1}B\left(D - C A^{-1}B\right)^{-1}
    \\
    -\left(D - C A^{-1}B\right)^{-1}CA^{-1} & \left(D - C A^{-1}B\right)^{-1}
    \end{pmatrix}
   \nonumber
   \\
   &=
   \begin{pmatrix}
   A^{-1}&0
    \\
   0 & 0
    \end{pmatrix}
    +
   \left(D - C A^{-1}B\right)^{-1} \begin{pmatrix}
  A^{-1}B CA^{-1}&-A^{-1}B
    \\
    -CA^{-1} & 1
    \end{pmatrix}\,.
\end{align}
As before, $[A^{-1}]_{jk} = \Theta(j - k)e^{i(j - k)\theta}$ and
\begin{align}
    \left[CA^{-1}B\right]_{jk} &= 
    \sum_{lm} C_{jl}A^{-1}_{lm}B_{mk} =
    \sum_{lm} (-e^{i\theta} \delta_{j,1}\delta_{l,N-1})A^{-1}_{lm}(-e^{-i(N - 1)\theta}e^{-\beta\hbar\Omega_s}\delta_{m,1}\delta_{k,1})
   \nonumber
   \\
   &=
   e^{i\theta} \delta_{j,1}A^{-1}_{N-1,1}e^{-i(N - 1)\theta}e^{-\beta\hbar\Omega_s}\delta_{k,1}
   =
  e^{i\theta} \delta_{j,1}e^{i(N-2)\theta}e^{-i(N - 1)\theta}e^{-\beta\hbar\Omega_s}\delta_{k,1}
   =
  \delta_{j,1}e^{-\beta\hbar\Omega_s}\delta_{k,1}
  \,,
   \\
   \left[A^{-1}B\right]_{jk} &= 
    \sum_{m} A^{-1}_{jm}B_{mk} =
    \sum_{m} A^{-1}_{jm}(-e^{-\beta\hbar\Omega_s}
    e^{-i(N - 1)\theta}\delta_{m,1}\delta_{k,1})
    =
   -A^{-1}_{j1}e^{-\beta\hbar\Omega_s}e^{-i(N - 1)\theta}\delta_{k,1}
   \nonumber
   \\
   &=
   -\Theta(j - 1)e^{i(j - 1)\theta}e^{-\beta\hbar\Omega_s}e^{-i(N - 1)\theta}\delta_{k,1}
   =
   -e^{-\beta\hbar\Omega_s}e^{-i(N-j)\theta}\delta_{k,1}\,,
   \\
   \left[CA^{-1}\right]_{jk} &= 
    \sum_{l} C_{jl}A^{-1}_{lk} =
    \sum_{l} (-e^{i\theta} \delta_{j,1}\delta_{l,N-1})A^{-1}_{lk}
    =
    -e^{i\theta}\delta_{j,1}A^{-1}_{N-1,k}
    \nonumber
    \\
   &  =
    -e^{i\theta}\delta_{j,1} \Theta(N-1 - k)e^{i(N-1 - k)\theta}
    =
    -\delta_{j,1}e^{i(N - k)\theta}\,,
\end{align}
leading to $\left(D - C A^{-1}B\right)^{-1} = (1 - e^{-\beta\hbar\Omega_s})^{-1} = n_B(\Omega_s) + 1$. In addition,
\begin{align}
    \left[ A^{-1}B CA^{-1}\right]_{jk}& = \sum_l
    \left[A^{-1}B\right]_{jl}
    \left[CA^{-1}\right]_{lk}
    =
    \sum_l
    A^{-1}_{j1}e^{-\beta\hbar\Omega_s}e^{-i(N - 1)\theta}\delta_{l,1}
    e^{i\theta}\delta_{l,1}A^{-1}_{N-1,k}
    \nonumber
    \\
    &=
   \Theta( j-1)e^{i(j - 1)\theta}e^{-\beta\hbar\Omega_s}e^{-i(N - 1)\theta}
    e^{i\theta}\Theta(N-1 - k)e^{i(N-1 - k)\theta}
    \nonumber
    \\
    &=
   e^{-\beta\hbar\Omega_s}
   e^{i(j-k)\theta}\,,
\end{align}
which yields
\begin{align}
   \left[ A^{-1} + \left(D - C A^{-1}B\right)^{-1} A^{-1}B CA^{-1}\right]_{jk} &= 
   \Theta(j - k)e^{i(j - k)\theta}
   +
   \left[n_B(\Omega_s) + 1\right]e^{-\beta\hbar\Omega_s}e^{i(j-k)\theta}
   \nonumber
   \\
    &= 
   \left[\Theta(j - k) 
   +
   \left[n_B(\Omega_s) + 1\right]e^{-\beta\hbar\Omega_s}\right]e^{i(j-k)\theta}
   \nonumber
   \\
    &= 
   \left[\Theta(j - k)
   +
  n_B(\Omega_s)\right]e^{i(j-k)\theta}\,.
\end{align}
One can see that the same form holds for the remaining terms of $\left(d - c a^{-1}b\right)^{-1}$, in agreement with $G^{--}_s$ for $\theta = \Delta\Omega_s$.

Having obtained $\left(d - c a^{-1}b\right)^{-1}$, we can calculate the remaining three quadrants of $Y^{-1}$ as follows:
\begin{align}
    \left[-a^{-1}b\left(d - c a^{-1}b\right)^{-1}\right]_{jk} &= \sum_l\left[-a^{-1}b\right]_{jl}\left[\left(d - c a^{-1}b\right)^{-1}\right]_{lk} 
    \nonumber
    \\
    &= 
    \sum_l a^{-1}_{j1}e^{-\beta\hbar\Omega_s}\delta_{l,N}
   \left[\Theta(l - k)
   +
  n_B(\Omega_s)\right]e^{i(l-k)\theta}
   \nonumber
    \\
    &= 
    e^{-\beta\hbar\Omega_s}
   \left[1
   +
  n_B(\Omega_s)\right]e^{i(N-k-j+1)\theta}
  = 
   n_B(\Omega_s)e^{i(N-k-j+1)\theta}\,,
   \\
   \left[ -\left(d - c a^{-1}b\right)^{-1}ca^{-1} \right]_{jk} &=
   \sum_l
   \left[ -\left(d - c a^{-1}b\right)^{-1}\right]_{jl}
   \left[ca^{-1} \right]_{lk}
   \nonumber
   \\
   &=
   \sum_l
   \left[\Theta(j - l)
   +
  n_B(\Omega_s)\right]e^{i(j-l)\theta}
    \delta_{l,1}a^{-1}_{Nk}
    \nonumber
   \\
   &=
   \left[1
   +
  n_B(\Omega_s)\right]
   e^{-i(N - k-j+1)\theta}\,,
   \\
   \left[a^{-1} +a^{-1}b \left(d - c a^{-1}b\right)^{-1}ca^{-1}\right]_{jk} &=
   \left[a^{-1}\right]_{jk}
   +
   \sum_{lm}\left[a^{-1}b \right]_{jl}
   \left[\left(d - c a^{-1}b\right)^{-1}\right]_{lm}
   \left[ca^{-1}\right]_{mk}
   \nonumber
   \\
   &=
   \Theta(j - k)e^{-i(j - k)\theta}
   +
   \sum_{lm}
   a^{-1}_{j1}e^{-\beta\hbar\Omega_s}\delta_{l,N} 
    \left[\Theta(l - m) 
   +
  n_B(\Omega_s)\right]e^{i(l-m)\theta}
   \delta_{m,1}a^{-1}_{Nk}
   \nonumber
   \\
   &=
   \Theta(j - k)e^{-i(j - k)\theta}
   +
   e^{-i(j - 1)\theta}e^{-\beta\hbar\Omega_s}
    \left[1
   +
  n_B(\Omega_s)\right]e^{i(N-1)\theta}
    e^{-i(N - k)\theta}
    \nonumber
   \\
   &=
   \Theta(j - k)e^{-i(j - k)\theta}
   +
   e^{-i(j-k)\theta}e^{-\beta\hbar\Omega_s}
    \left[1
   +
  n_B(\Omega_s)\right]
   \nonumber
   \\
   &=
   \left[\Theta(j-k) + n_B(\Omega_s)\right]e^{-i(j - k)\theta}\,.
\end{align}

\section{Correlation Tensor}
\label{sec:Correlation_Tensor}

\begin{align}
    \langle \tilde{\mathbf{f}}_n\otimes\tilde{\mathbf{f}}_l\rangle
    &=
    \left\langle\sum_s 
     \nabla Y_s\left(\mathbf{r}_n\right)
     \left[
   e^{i \Delta\Omega_s n}
\xi_s
     +
   e^{-i\Delta \Omega_s n }
    \xi^*_s
    \right]\otimes
    \sum_{s '}
     \nabla Y_{s'}\left(\mathbf{r}_l\right)
     \left[
   e^{i \Delta\Omega_{s'}l}
\xi_{s'}
     +
   e^{-i \Delta\Omega_{s'} l }
    \xi^*_{s'}
    \right]\right\rangle
    \nonumber
    \\
    &=
    \int \prod_{s''} d\xi_{s''}d\xi_{s''}^*
    \sum_s 
     \nabla Y_s\left(\mathbf{r}_n\right)
     \left[
   e^{i \Delta\Omega_s n}
\xi_s
     +
   e^{-i\Delta \Omega_s n }
    \xi^*_s
    \right]
    \nonumber
\otimes
    \sum_{s '}
     \nabla Y_{s'}\left(\mathbf{r}_l\right)
     \left[
   e^{i \Delta\Omega_{s'}l}
\xi_{s'}
     +
   e^{-i\Delta \Omega_{s'} l }
    \xi^*_{s'}
    \right]P_{s''}
    \nonumber
    \\
    &=
    \sum_s 
    \int d\xi_{s}d\xi_{s}^*
     \nabla Y_s\left(\mathbf{r}_n\right)
     \left[
   e^{i \Delta\Omega_s n}
\xi_s
     +
   e^{-i \Delta \Omega_s n }
    \xi^*_s
    \right]
    \otimes
     \nabla Y_s\left(\mathbf{r}_l\right)
     \left[
   e^{i \Delta\Omega_{s}l}
\xi_{s}
     +
   e^{-i\Delta \Omega_{s} l }
    \xi^*_{s}
    \right]P_{s}
    \nonumber
    \\
    &=
    \sum_s 
     \nabla Y_s\left(\mathbf{r}_n\right)
     \otimes
     \nabla Y_s\left(\mathbf{r}_l\right)
    \int d\xi_{s}d\xi_{s}^*
     \left[
   e^{i \Delta\Omega_sn}
\xi_s
     +
   e^{-i \Delta\Omega_s n }
    \xi^*_s
    \right]
     \left[
   e^{i \Delta\Omega_{s}l}
\xi_{s}
     +
   e^{-i \Delta\Omega_{s} l }
    \xi^*_{s}
    \right]P_{s}
    \nonumber
    \\
    &=
    \sum_s 
     \nabla Y_s\left(\mathbf{r}_n\right)
     \otimes
     \nabla Y_s\left(\mathbf{r}_l\right)
     \langle\xi_s\xi_s^*\rangle
    2\cos\left[\Delta\Omega_s\left(n - l\right)\right]
    \nonumber
    \\
    &=
    \sum_s 
     \nabla Y_s\left(\mathbf{r}_n\right)
     \otimes
     \nabla Y_s\left(\mathbf{r}_l\right)
     \coth\left(\frac{\beta\hbar\Omega_s}{2}\right)\cos\left[\Delta \Omega_s\left(n - l\right)\right]\,.
\end{align}

\section{Recoil Term}
\label{sec:Recoil_Term}

\begin{align}
    &\sum_{l = 1}^n
     \sin\left[\Delta\Omega_s\left(n  - l\right)\right] Y_s(\mathbf{r}_l)
     \nonumber
     \\
     =&
   \sum_{l = 1}^n \frac{\cos\left[\Delta\Omega_s\left(n  - l\right)\right]
    \cos\left[\Delta\Omega_s\right]-\cos\left[\Delta\Omega_s\left(n + 1 - l\right)\right]}{\sin\left[\Delta\Omega_s\right]} Y_s(\mathbf{r}_l)
    \nonumber
     \\
     =&
   \sum_{l = 1}^n \frac{\cos\left[\Delta\Omega_s\left(n  - l\right)\right]
    \cos\left[\Delta\Omega_s\right]}{\sin\left[\Delta\Omega_s\right]} Y_s(\mathbf{r}_l)
    -
    \sum_{l = 0}^{n-1} \frac{\cos\left[\Delta\Omega_s\left(n  - l\right)\right]}{\sin\left[\Delta\Omega_s\right]} Y_s(\mathbf{r}_{l+1})
     \nonumber
     \\
     =&
   \sum_{l = 1}^{n-1} \frac{\cos\left[\Delta\Omega_s\left(n  - l\right)\right]
    \cos\left[\Delta\Omega_s\right]}{\sin\left[\Delta\Omega_s\right]} Y_s(\mathbf{r}_l)
    -
    \sum_{l = 1}^{n-1} \frac{\cos\left[\Delta\Omega_s\left(n  - l\right)\right]}{\sin\left[\Delta\Omega_s\right]} Y_s(\mathbf{r}_{l+1})
    +
   \frac{
    \cos\left[\Delta\Omega_s\right]}{\sin\left[\Delta\Omega_s\right]} Y_s(\mathbf{r}_n)
    -
    \frac{\cos\left[\Delta\Omega_s n\right]}{\sin\left[\Delta\Omega_s\right]} Y_s(\mathbf{r}_{1})
    \nonumber
    \\
     =&
     -
    \sum_{l = 1}^{n-1} \frac{\cos\left[\Delta\Omega_s\left(n  - l\right)\right]}{\sin\left[\Delta\Omega_s\right]} \left[Y_s(\mathbf{r}_{l+1})-Y_s(\mathbf{r}_l)\right]
    +
   \frac{
    \cos\left[\Delta\Omega_s\right]}{\sin\left[\Delta\Omega_s\right]} Y_s(\mathbf{r}_n)
    -
    \frac{\cos\left[\Delta\Omega_s n\right]}{\sin\left[\Delta\Omega_s\right]} Y_s(\mathbf{r}_{1})
    \nonumber
    \\
     \approx&
     -
    \sum_{l = 1}^{n-1} \frac{\cos\left[\Delta\Omega_s\left(n  - l\right)\right]}{\sin\left[\Delta\Omega_s\right]} 
    \nabla_\mathbf{r}Y_s(\mathbf{r}_{l})
    \cdot\left(\mathbf{r}_{l+1}-\mathbf{r}_{l}\right)
    +
   \frac{
    \cos\left[\Delta\Omega_s\right]}{\sin\left[\Delta\Omega_s\right]} Y_s(\mathbf{r}_n)
    -
    \frac{\cos\left[\Delta\Omega_s n\right]}{\sin\left[\Delta\Omega_s\right]} Y_s(\mathbf{r}_{1})
    \nonumber
    \\
     \approx&
     -
    \sum_{l = 1}^{n-1} \frac{\cos\left[\Delta\Omega_s\left(n  - l\right)\right]}{\Omega_s} 
    \nabla_\mathbf{r}Y_s(\mathbf{r}_{l})
    \cdot\dot{\mathbf{r}}_l
    +
   \frac{
    \cos\left[\Delta\Omega_s\right]}{\Delta\Omega_s} Y_s(\mathbf{r}_n)
    -
    \frac{\cos\left[\Delta\Omega_s n\right]}{\Delta\Omega_s} Y_s(\mathbf{r}_{1})\,,
\end{align}
where the last expression holds in the $\Delta\rightarrow0$ limit.

\twocolumngrid

\end{document}